# Universal metasurfaces for complete linear control of coherent light transmission

*Taeyong Chang*[†], *Joonkyo Jung*[†], *Sang-Hyeon Nam, Hyeonhee Kim, Jong Uk Kim, Nayoung Kim, Suwan Jeon, Minsung Heo,* and *Jonghwa Shin*[*]

[†]These authors equally contributed to this work.

Taeyong Chang, Joonkyo Jung, Sang-Hyeon Nam, Hyeonhee Kim, Jong Uk Kim, Nayoung Kim, Suwan Jeon, Minsung Heo, and Jonghwa Shin

Department of Materials Science and Engineering, KAIST,

291 Daehak-ro, Yuseong-gu, Daejeon 34141, Republic of Korea

E-mail (corresponding author): qubit@kaist.ac.kr

**All co-authors E-mail:**

Taeyong Chang: taeyong08@kaist.ac.kr

Joonkyo Jung: kyo2531@kaist.ac.kr

Sang-Hyeon Nam: indigo7@kaist.ac.kr

Hyeonhee Kim: khh0106l@kaist.ac.kr

Jong Uk Kim: jonguk.jay.kim@gmail.com

Nayoung Kim: nykim11@kaist.ac.kr

Suwan Jeon: jswoo325@kaist.ac.kr

Minsung Heo: heo0324@gmail.com

Jonghwa Shin (Corresponding Author): qubit@kaist.ac.kr

**Acknowledgments:** The authors thank Michael H. Hwang (SEMCRON) for the highly reliable electron beam lithography. This work is supported by National Research Foundation of Korea

(NRF) grants funded by the Korea government (MSIT) (NRF-2018M3D1A105899821, NRF-2021R1A2C200868711), and a KAIST Grand Challenge 30 Project (KC30) grant funded by KAIST (N11210118).

**Competing interests:** The authors declare no competing interests.

**Data availability statement:** The data that support the findings of this study are available from the corresponding author upon reasonable request.

**Author contributions:** T.C. conceived the idea and J.S. supervised the project. T.C., J.J., J.S., N.K., S.J., and M.H. conducted the theoretical analyses. J.J. performed the numerical simulation. T.C., J.J., S.N., and J.K. fabricated the samples. T.C., J.J., and H.K. characterized them optically. J.J., T.C., and J.S. prepared the manuscript.

## Abstract

Recent advances in metasurfaces and optical nanostructures have enabled complex control of incident light with optically thin devices. However, it has thus far been unclear whether it is possible to achieve complete linear control of coherent light transmission, i.e., independent control of polarization, amplitude, and phase for both input polarization states, with just a single, thin nanostructure array. Here we prove that it is possible and propose a universal metasurface, a bilayer array of high-index elliptic cylinders, that possesses a complete degree of optical freedom with fully designable chirality and anisotropy. We mathematically show the completeness of achievable light control with corresponding Jones matrices, experimentally demonstrate new types of three-dimensional holographic schemes that were formerly impossible, and present a systematic way of realizing any input-state-sensitive vector linear optical device. Our results unlock previously inaccessible degrees of freedom in light transmission control.

## 1. Introduction

Polarization, intensity, and phase are fundamental properties of coherent light, and controlling them with a high degree of freedom is the central objective for numerous optical devices. Controlling all three properties independently for a single input polarization state is a difficult task in itself, but doing so for all possible input polarizations in a precisely designed manner is the ultimate goal: it allows the complete control of coherent light transmission, fully utilizing the vector nature of light. Historically, birefringent materials such as calcites have been used for polarization-dependent light control[1,2]. Although various crystal classes of natural optical materials possess anisotropy, chirality, or both, the effects of these characteristics are not strong enough for a thin slab to achieve the full range of controllability of the optical response. Also, such materials are usually spatially homogeneous and do not allow easy control of transmitted amplitudes[1]. Recent advanced metasurfaces are reanimating the field with their ability to control light at a pixel-by-pixel spatial degree of freedom, allowing novel applications ranging from imaging[3,4] and communications[5,6] to quantum optical operations[7,8]. However, the currently achievable range of light transmission control from a single metasurface is still limited. This limitation translates into restricted accessible sets of Jones matrices, which can quantitatively represent thin optical media in the paraxial regime[9,10]. In particular, conventional waveplates are restricted to unitary symmetric Jones matrices under linear polarization basis, i.e., they are achiral. This limitation remains the same for many metasurfaces including even very recently developed[11-19]. In response, there have been active efforts to surmount this restriction and realize all possible Jones matrices, enabling bespoke control of polarization, amplitude, and phase, i.e., achieving complete control of light transmission.

First, gaining chirality and breaking the symmetry of the Jones matrix has been attempted by removing the mirror symmetry in the spatial configuration, for example via utilization of three-dimensional (3D) chiral (meta-)atoms[20], cascades of different waveplates[1] or metasurfaces[21-26], and off-normal incidence[27,28]. Second, the unitary condition, which is associated with the conservation of energy between input and output beams, can be avoided by utilizing gain materials[29] or incorporating controllable radiative loss realized by superposition[30] or transversal homogenization[18,19,31-33] of two output states with different phases. However, all the above cases still achieve only subsets of the complete Jones matrix set. Prior extents of pixel-by-pixel controllability of transmitted coherent light are compared in the Venn diagram in Fig. 1A. We note that, while there have been significant advancements in realizing symmetric Jones matrices, a vast area related to asymmetric Jones matrices is still unreachable, preventing full designability of chirality and achievement of a full degree of freedom of coherent light transmission control.

Here, we propose a thin effective optical material composed of a bilayer array of dielectric nanostructures, a universal metasurface, and theoretically and experimentally demonstrate that such a metasurface can embody, at ~2 μm thickness, any arbitrary passive Jones matrix array and gain pixel-wise complete linear control of coherent light transmission. For such a universal metasurface, the achievable optical control is expanded to the full set as shown in Fig. 1A. The theoretical proof is based on a recent report in the field of mathematics on the factorization of an arbitrary unitary matrix[34]. We experimentally demonstrate a new holographic scheme that is impossible with an existing optical medium or metasurface. Furthermore, we theoretically prove that any arbitrary passive vector linear optical device can be constructed systematically by using only such universal metasurfaces and conventional lenses. As an example, using only three universal metasurfaces and two lens arrays, we numerically demonstrate parallelized probabilistic linear quantum optical controlled-NOT (CNOT) gates for an array of two single-photon qubits[2,35], whose individual spatial channel is functionally equivalent to the CNOT gate implemented previously by bulk optics[35]. The universal metasurfaces provide a full degree of freedom of coherent light transmission control by a thin optical component, and can serve as a new optical platform.

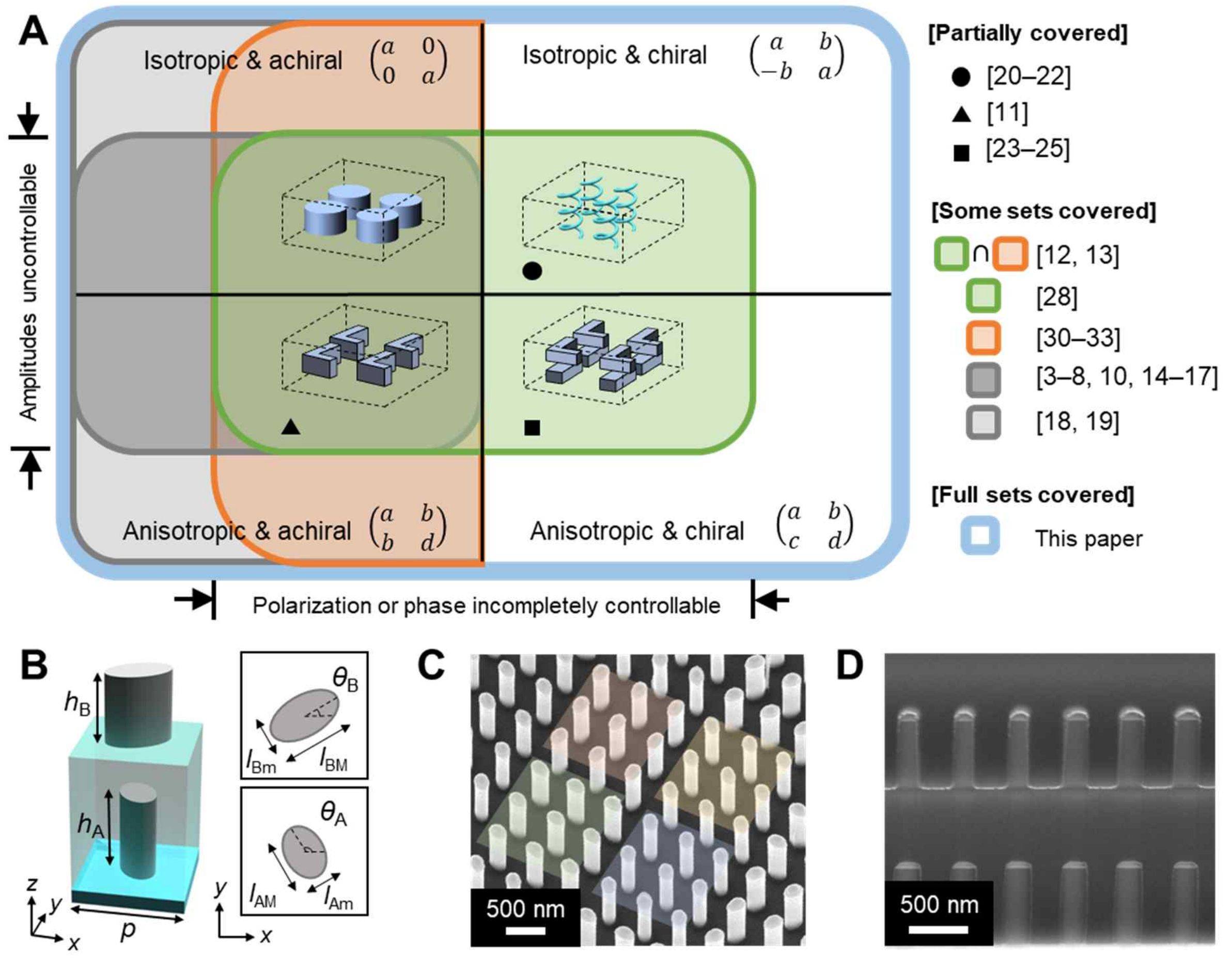

**Figure 1. Optical property and structure of universal metasurface.** (**A**) Venn diagram of transmitted coherent light controllability, each region of which has a set of nanostructures as an element. Representative metasurface nanostructures and corresponding Jones matrices are shown for each quadrant (*a*, *b*, *c*, and *d* are normalized complex numbers). Corners are rounded for visual clarity. (**B**) Schematic of unit structure of bilayer unitary metasurfaces. The top

(bottom) right panel shows horizontal cross section of layer B (layer A). $p$: period; $h$: height of cylinders; $l_{\mathrm{M,m}}$: length of the major (M) and minor (m) axes; $\theta$: orientation angle. (**C** to **D**) Scanning electron microscope images of a fabricated universal metasurface. A perspective view (**C**) and vertical cross section (**D**) of fabricated sample are shown. The recolored region represents a cluster of unit structures.

# 2. Results and Discussion

## 2.1. Theoretical Background and Design Rule for Universal Metasurfaces

We first prove that an arbitrary 2-by-2 unitary matrix can be decomposed into two unitary symmetric matrices, which is the key theoretical foundation for constructing universal metasurfaces. Since unitary symmetric matrices are orthogonally similar to unitary diagonal matrices, multiplication of two arbitrary 2-by-2 unitary symmetric matrices can be expressed as

$$U_{\mathrm{SB}}U_{\mathrm{SA}} = \begin{pmatrix} \cos\theta_{\mathrm{B}} & -\sin\theta_{\mathrm{B}} \\ \sin\theta_{\mathrm{B}} & \cos\theta_{\mathrm{B}} \end{pmatrix} \begin{pmatrix} e^{i\phi_{\mathrm{BM}}} & 0 \\ 0 & e^{i\phi_{\mathrm{Bm}}} \end{pmatrix} \begin{pmatrix} \cos\theta_{\mathrm{B}} & \sin\theta_{\mathrm{B}} \\ -\sin\theta_{\mathrm{B}} & \cos\theta_{\mathrm{B}} \end{pmatrix} \cdot \begin{pmatrix} \cos\theta_{\mathrm{A}} & -\sin\theta_{\mathrm{A}} \\ \sin\theta_{\mathrm{A}} & \cos\theta_{\mathrm{A}} \end{pmatrix} \begin{pmatrix} e^{i\phi_{\mathrm{AM}}} & 0 \\ 0 & e^{i\phi_{\mathrm{Am}}} \end{pmatrix} \begin{pmatrix} \cos\theta_{\mathrm{A}} & \sin\theta_{\mathrm{A}} \\ -\sin\theta_{\mathrm{A}} & \cos\theta_{\mathrm{A}} \end{pmatrix} \quad (1)$$

where $\theta_{\mathrm{A,B}}$ , $\phi_{\mathrm{AM,Am}}$ , $\phi_{\mathrm{BM,Bm}}$ $\in \mathbb{R}$. By choosing $\phi_{\mathrm{BM}} = 0$ and $\phi_{\mathrm{Bm}} = \pi$ , where each unitary symmetric matrix is parameterized by three real-valued angles: $\theta_X$, $\phi_{X\mathrm{M}}$, and $\phi_{X\mathrm{m}}$ ($X \in \{\mathrm{A, B}\}$). Since a 2-by-2 unitary matrix has four real degrees of freedom while the above parameterization has six real parameters, the mapping from two unitary symmetric matrices to one unitary matrix would be surjective. Thus, we choose $\phi_{\mathrm{BM}} = 0$ and $\phi_{\mathrm{Bm}} = \pi$ (i.e., a "half-wave retarder") to simplify the problem but note that it is not a unique way,

$$
\begin{aligned}
U_{\mathrm{SB}}U_{\mathrm{SA}} &= \begin{pmatrix} \cos 2\theta_{\mathrm{B}} & -\sin 2\theta_{\mathrm{B}} \\ \sin 2\theta_{\mathrm{B}} & \cos 2\theta_{\mathrm{B}} \end{pmatrix} \begin{pmatrix} 1 & 0 \\ 0 & -1 \end{pmatrix} \cdot \\
&\quad \begin{pmatrix} \cos\theta_{\mathrm{A}} & -\sin\theta_{\mathrm{A}} \\ \sin\theta_{\mathrm{A}} & \cos\theta_{\mathrm{A}} \end{pmatrix} \begin{pmatrix} e^{i\phi_{\mathrm{AM}}} & 0 \\ 0 & e^{i\phi_{\mathrm{Am}}} \end{pmatrix} \begin{pmatrix} \cos\theta_{\mathrm{A}} & \sin\theta_{\mathrm{A}} \\ -\sin\theta_{\mathrm{A}} & \cos\theta_{\mathrm{A}} \end{pmatrix} \\
&= \begin{pmatrix} \cos\theta'_{\mathrm{B}} & -\sin\theta'_{\mathrm{B}} \\ \sin\theta'_{\mathrm{B}} & \cos\theta'_{\mathrm{B}} \end{pmatrix} \begin{pmatrix} e^{i\phi_{\mathrm{AM}}} & 0 \\ 0 & e^{i(\phi_{\mathrm{Am}}+\pi)} \end{pmatrix} \begin{pmatrix} \cos\theta_{\mathrm{A}} & \sin\theta_{\mathrm{A}} \\ -\sin\theta_{\mathrm{A}} & \cos\theta_{\mathrm{A}} \end{pmatrix}
\end{aligned} \tag{2}
$$

where $\theta'_{\mathrm{B}} = 2\theta_{\mathrm{B}} - \theta_{\mathrm{A}}$. The last form of $U_{\mathrm{SB}}U_{\mathrm{SA}}$ in Eq. 2 is a general factorization of an arbitrary 2-by-2 unitary matrix[34] and reveals the bijective mapping between one 2-by-2 unitary matrix and two 2-by-2 unitary symmetric matrices under the (0, π)-phase constraint for one of the symmetric matrices■.[34] Physically, this suggests we can realize metasurfaces with arbitrary unitary Jones matrices (unitary metasurfaces) by forming a tandem structure made of two metasurfaces with arbitrary unitary symmetric Jones matrices (unitary symmetric metasurfaces). The schematic of unit structure of the unitary metasurface in Fig.1B shows the geometric parameters used as design variables. These structural parameters are directly related to the parameters in Eqs. 1–2 ($\theta_{\mathrm{A}}$, $\theta_{\mathrm{B}}$, $\phi_{\mathrm{AM}}$, $\phi_{\mathrm{Am}}$, $\phi_{\mathrm{BM}}$, and $\phi_{\mathrm{Bm}}$).

Once all unitary Jones matrices becomes accessible, the formation of universal metasurfaces with access to the full set of all possible Jones matrices including amplitude controllability is relatively straightforward: for example, one can utilize a transversal homogenization of unitary metasurfaces (i.e. introducing a super-cell composed of two unitary unit-cells)[19] such as $A = U\Sigma V^{\dagger} = \frac{1}{2}U(D_1 + D_2)V^{\dagger} = \frac{1}{2}(U_1 + U_2)$ for an arbitrary passive (i.e. the maximum singular value is one) Jones matrix $A$ where $U$, $V$, $U_1$, and $U_2$ are unitary matrices, $\Sigma$ is a diagonal matrix showing singular values, and $D_1$ and $D_2$ are unitary and diagonal matrices. In this scheme, the desired portion of the energy is reflected or deflected into higher diffraction orders other than the near-normal transmission that we are interested in (i.e. radiative losses)[18,31-33]. We note that a wider angular region of completely controllable transmission can be achieved with smaller unit structures of a universal metasurface. We fabricate unitary and universal metasurfaces targeting 915 nm wavelength by depositing silicon and dielectric encapsulation as shown in Figs. 1C–D (Methods).

The preceding discussion provide mathematical completeness of realizable arbitrary passive Jones matrix with universal metasurfaces. Now, we provide systematic design strategy for unitary metasurfaces and universal metasurfaces with a closed-form solution. For unitary metasurfaces, one can predict the output polarizations and phases based on such structural parameters by first reformulating the Jones matrix with input-output polarization pairs and corresponding phase advances. Without loss of generality, we can represent an arbitrary unitary Jones matrix in the form $Q = WPC^{\dagger} = [|w\rangle,\ |w_{\perp}\rangle][e^{i\phi_1},\ 0;\ 0,\ e^{i\phi_2}][|R\rangle,\ |L\rangle]^{\dagger}$ where $|R\rangle = [1;\ i]/\sqrt{2}$ (right-handed circularly polarized state, RCP) and $|L\rangle = [1;\ -i]/\sqrt{2}$ (left-handed circularly polarized state, LCP) are used as a basis for expressing input polarization states and corresponding output polarization states are denoted with $|w\rangle$ and $|w_{\perp}\rangle$. We now find an analytic relationship between the geometric parameters ($\theta_{\mathrm{A}}$, $\theta_{\mathrm{B}}$, $\phi_{\mathrm{AM}}$, and $\phi_{\mathrm{Am}}$) and output polarization states and phases ($2\psi$, $2\chi$, $\phi_1$, and $\phi_2$ where $2\psi$ and $2\chi$ are the spherical coordinates of $|w\rangle$ on the Poincaré sphere). $Q$ can also be represented as

$$Q = \begin{pmatrix} \cos\psi & -\sin\psi \\ \sin\psi & \cos\psi \end{pmatrix} \begin{pmatrix} \cos\chi & \sin\chi \\ i\sin\chi & -i\cos\chi \end{pmatrix} \begin{pmatrix} e^{-i\phi_1'} & 0 \\ 0 & e^{-i\phi_2'} \end{pmatrix} \begin{pmatrix} 1/\sqrt{2} & -i/\sqrt{2} \\ 1/\sqrt{2} & i/\sqrt{2} \end{pmatrix} \tag{3}$$

where $\phi_i' = \phi_i - \phi_i''$ ($i = 1,2$), $\phi_1'' = \angle(\cos\psi\cos\chi - i\sin\psi\sin\chi)$ and $\phi_2'' = \angle(\cos\psi\sin\chi + i\sin\psi\cos\chi)$. By equating Eq. 2 and Eq. 3, analytic relations between design parameters and output polarization including phases can be obtained (Supplementary text S1):

$$\begin{cases} \theta_{\mathrm{A}} = \dfrac{\phi_1' - \phi_2'}{2} + \dfrac{\pi}{4}, \\ \theta_{\mathrm{B}} = \dfrac{\psi}{2} + \dfrac{\phi_1' - \phi_2'}{4} + \dfrac{\pi}{4}, \\ \phi_{\mathrm{AM}} = \dfrac{\phi_1' + \phi_2'}{2} + \chi + \dfrac{7\pi}{4}, \\ \phi_{\mathrm{Am}} = \dfrac{\phi_1' + \phi_2'}{2} - \chi + \dfrac{5\pi}{4}. \end{cases} \tag{4}$$

We designed unitary metasurfaces for target Jones matrices based on this solution. We note that the resulting output polarization states and phases are regularly spaced for regular arrays of geometric parameters under the incidence of circularly polarized states, without any abrupt changes. This simplifies the design process and also hints at relative robustness of the design against possible imperfection in fabrication. In addition, even though the analytic solution in Eq. 4 is expressed for circular polarization incidence ($Q = WPC^{\dagger}$), one can use it for any desired input polarization pair ( $E^{\dagger} = [|e\rangle,\ |e_{\perp}\rangle]^{\dagger}$ ) with any desired target output polarization pair ( $W_E$ ) and phase pair ( $P_E$ ) such that $Q_E = W_E P_E E^{\dagger} = W_E P_E E^{\dagger} C C^{\dagger} = W_C P_C C^{\dagger}$ where $W_C$ , $P_C$ can be calculated from $W_E P_E E^{\dagger} C$ .

For universal metasurfaces of arbitrary Jones matrix $A$ for any desired input polarization pair with any desired target output polarization pair, amplitude pair, and phase pair (subject to passiveness), one can find $U_1$ and $U_2$ such that $A = \frac{1}{2}(U_1 + U_2)$ and apply above design strategy to find all geometric parameters for the realization. For simplicity of the proof-of-concept, we composed square clusters with 3 to 15 unit structures in each direction to minimize the change in the optical properties of unit structures due to near-field coupling in heterogeneous configurations. One can further reduce the size of clusters by optimizing the shape of each elliptic cylinder to compensate for near-field effects.

**2.2. Numerical Validations for Unitary Metasurfaces and Universal Metasurfaces**

Finite-difference time-domain (FDTD) simulations were employed to confirm the generality of the proposed unitary metasurfaces (Fig. 2). Based on Eq. 3, output polarization states and phase delays under two orthogonal input polarization states (e.g. RCP and LCP) for a unitary metasurface can be represented as $|w\rangle$ , $\phi_1$ and $|w_{\perp}\rangle$ , $\phi_2$ [14,15]. Using unitary metasurfaces with proper geometric parameters, simulation results show that transmission phases can be independently controlled over the entire phase space for the incidence of (without loss of generality) RCP and LCP (Fig. 2A), while the output polarization states are intentionally kept constant for all cases (Fig. 2B) (Methods). Note that the target output polarization states were arbitrarily chosen by a random number generator to demonstrate the generality of the result. This is a prominent difference between unitary and unitary symmetric metasurfaces and reveals the continuous controllability of chirality by general unitary metasurfaces. There is no set of unitary symmetric Jones matrices that can provide an arbitrary pair of transmission phases for an arbitrary unitary conversion of polarization states except when output states are exact conjugations of input states[14,15]. However, input-output polarization pairs (e.g., RCP- $|w\rangle$ and LCP- $|w_{\perp}\rangle$ ) of unitary metasurfaces have no such restriction. Moreover, the range of achievable chirality in this study is sufficient to realize a full degree of freedom in coherent light transmission control, and fills all previously unoccupied sets.

To show the arbitrary controllability of a Jones matrix of a universal metasurface, we first randomly generate a Jones matrix and implement it with a universal metasurface (Fig. 2C). We then arbitrarily shift one individual matrix component of the given Jones matrix in Fig. 2C while keeping the other components constant and try to find universal metasurfaces for the resulting Jones matrices as well (left and right panels in Fig. 2D). For all cases, the target Jones matrices and those retrieved from the FDTD simulation of the corresponding universal metasurfaces are well matched (Methods). These are randomly generated examples and illustrate the generality of the proposed universal metasurface.

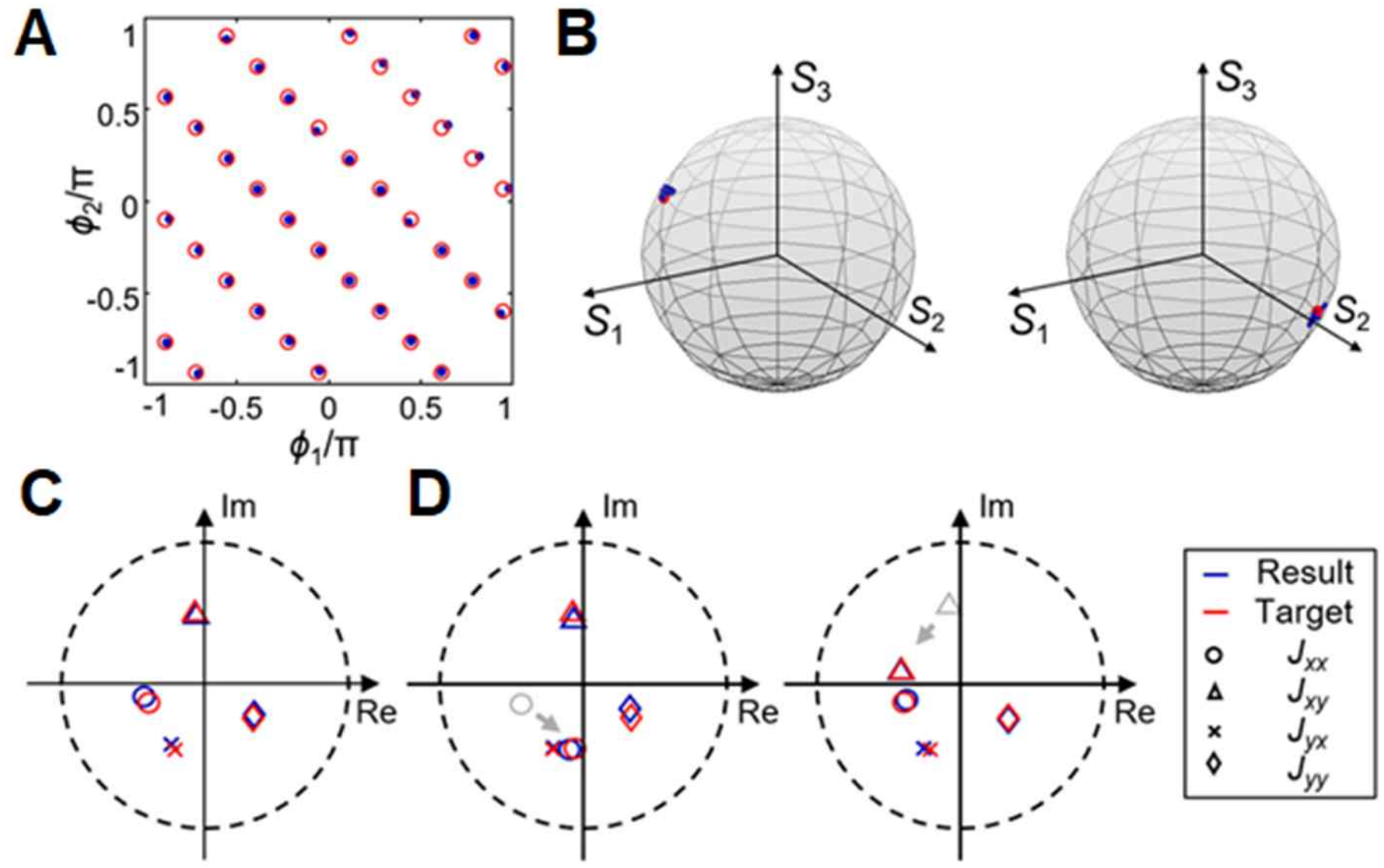

**Figure 2. Numerical validation of unitary and universal metasurface.** (**A**–**B**) Arbitrary wave retardation for two arbitrary orthogonal input-output polarization pairs based on a full degree of freedom of chirality and anisotropy of unitary metasurfaces. (**A**) Transmission phases for RCP (LCP) input, $\phi_1$ ($\phi_2$). Red circles indicate target phases. Blue dots are phases retrieved from FDTD simulations. (**B**) Output polarization states for RCP (left panel) and LCP (right panel) inputs on the Poincaré sphere. Red (Blue) dot shows target (retrieved) polarization states. (**C** to **D**) Elements of the Jones matrices on a complex plane. Gray markers show original elements being shifted. Dashed lines indicate unit magnitude.

## 2.3. Demonstration of Complete Linear Control of Coherent Light Transmission

We experimentally demonstrate complete vectorial wavefront manipulation enabled by unitary and universal metasurfaces through four examples (Figs. 3 and 4). Throughout this paper, we visualize polarization states and intensity measured with full-stokes polarimetry (Fig. S1) with false colors by directly mapping the Poincaré sphere into the CIELAB color space (Fig. 3A). For example, red, green, white, and black indicate $x$-, $y$-polarized, RCP, and LCP states, respectively.

For a unitary metasurface with spatially varying structures, output polarization states with different phase delays under two orthogonal input polarization states (e.g. RCP and LCP) can be represented as spatially dependent forms, $|w\rangle(\mathbf{r})$, $\phi_1(\mathbf{r})$ and $|w_\perp\rangle(\mathbf{r})$, $\phi_2(\mathbf{r})$, respectively. With the full set of unitary Jones matrices available, one can arbitrarily design the output polarization state pairs in a pixel-by-pixel manner (i.e., vector profiles), with independent control of the phase delay for each output polarization (also in a pixel-wise manner), which was impossible for previous metasurfaces. Figure 3B shows two orthogonal Poincaré beams[36] generated with independently controlled phase profiles under RCP and LCP Gaussian beams illumination (Methods). The demonstration in Fig. 3B directly contrasts with the example in Ref. 14, which shows correlated phase profiles for two orthogonal output vector beams. In a similar manner, we show two independent holographic images whose polarization states are not conjugations of corresponding input states (Fig. 3C), which is in direct contrast to Ref. 15.

Universal metasurfaces can provide even more generalized vectorial wavefront manipulation. In Fig. 3D, we show two independent vectorial holographic images under two orthogonal input states that require arbitrary controllable asymmetric and non-unitary Jones matrices (Methods). This is in contrast to Ref. 16,17,37,38, which show the generations of vectorial holographic images for a single input polarization state, and Ref. 10, which can be interpreted as providing vectorial holographic images for different input polarization states that cannot be arbitrary and are not independent of each other.

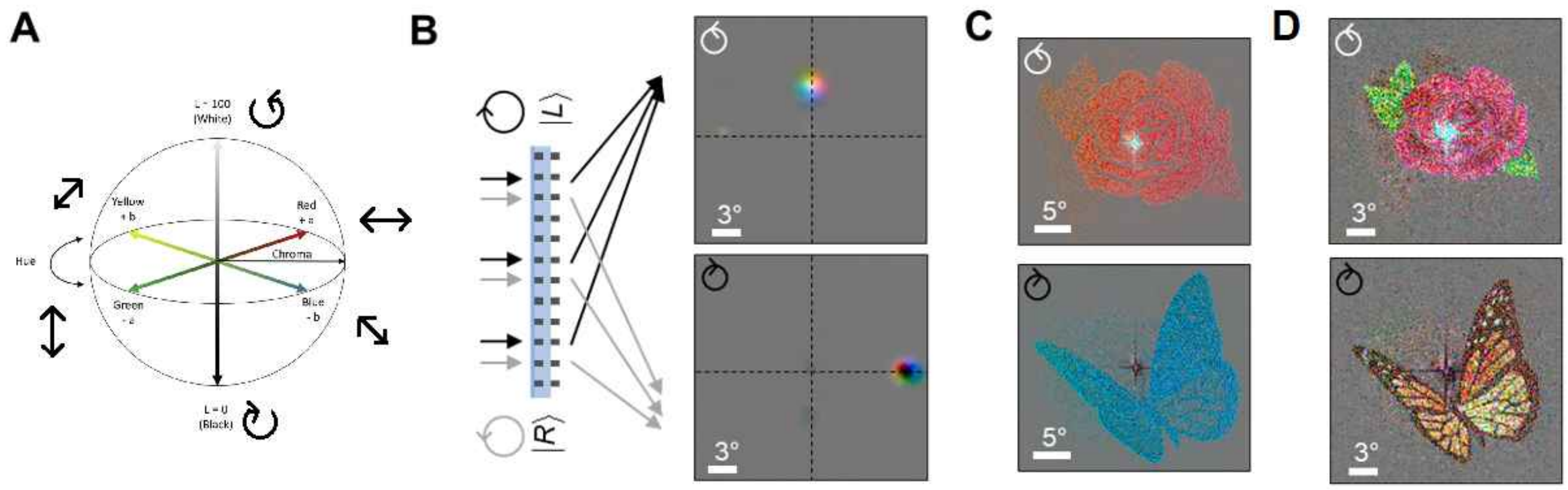

**Figure 3. Experimental demonstration of complete vectorial wavefront manipulation.** (**A**) Three-dimensional mapping of CIELAB color space on the Poincaré sphere where the conversion relation is given as ($L^*$, $a^*$, $b^*$) = ($50\cdot(S_3+1)$, $100\cdot S_1$, $100\cdot S_2$). Intensity, the orientation of polarization states, and the ellipticity of polarization states are described as chroma, hue, and lightness, respectively. (**B**) Two Poincaré beam generations with independent phase profiles (a schematic in left panel). Target values of deflection angle and divergence angle of a Poincaré beam in left (right) panel are 5° along $+y$ and 1.5° (10° along $+x$ and 0.8°), respectively. Input polarization states are denoted within each panel. (**C**) Two independent holographic images. (**D**) Two independent vectorial holographic images.

The most general type of vectorial wavefront manipulation based on a universal metasurface is illustrated in Fig. 4. Based on the electromagnetic equivalence principle, an arbitrary set of sources inside a closed surface can be replaced with equivalent surface sources with exactly the same resulting fields in the outside volume. Hence, if one can realize, with a single metasurface, any combination of two completely unrelated arbitrary tangential field profiles on the transmission-side of an infinite plane for two incident uniform plane waves with different polarization states, this means that it is possible to obtain any combination of two unrelated 3D electromagnetic field profiles on the transmission-side space that would be obtainable by two sets of virtual sources in the incident-side semi-infinite space, using just the given metasurface and two plane waves (Fig. 4A–B). This is different from typical volumetric holography because the phase and polarization profiles in addition to the intensity profiles, can be exactly designed, leading to potential for new applications. As an illustrative example, we designed a universal metasurface that produces a 3D spiral pattern with radial polarization under RCP incidence and produces the letters “KAI” and “ST” on different transverse planes with gradually varying polarization states following the circumference of the $S_1$-$S_3$ plane of a Poincaré sphere under LCP incidence (Fig. 4C) (Methods). Experimental results show good agreement with numerical predictions (Fig. 4D, Supplementary Fig. S2). We note that the result in Fig. 4D is enabled by the arbitrary polarization, amplitude, and phase controllability of the universal metasurface. In particular, it is revealed that both full control of the target 3D holographic patterns and design of their point-by-point polarization states are possible, in addition to control of their intensity and phase patterns, all independently for the two polarization states of the incident beam, which was previously unheard-of for a single metasurface. The demonstration in Fig. 4D contrasts with Ref. 39, which shows a single scalar (spatially homogeneous polarization state) 3D hologram, Ref. 38, which show a 3D vectorial hologram with a single input polarization state, and Ref. 19, which can be interpreted as providing 3D vectorial holograms for different input polarization states that cannot be arbitrary and are not independent of each other. We compare all experimental data in Figs. 3–4 with numerical calculations in Supplementary Fig. S2 and show measured raw data in Supplementary Fig. S3.

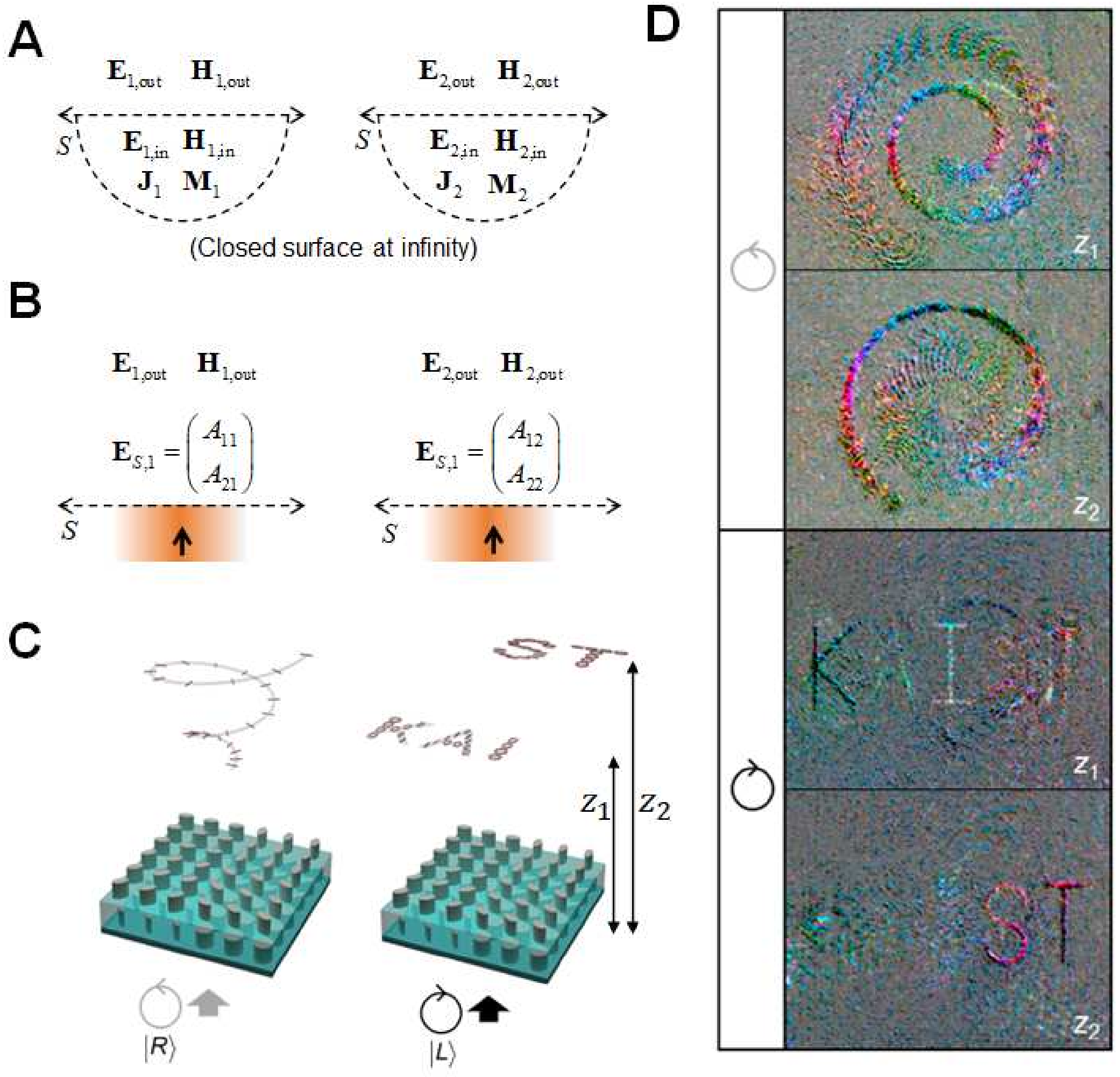

**Figure 4. Experimental demonstrations of two independent 3D vector field profiles.** (**A**) Two solutions of Maxwell's equation in a source-free volumetric region outside the enclosed surface, *S*. (**B**) Two electromagnetic field profiles outside *S* remain unchanged by realizing the same tangential electric fields at *S*, which can be done by a universal metasurface under illuminations of two plane waves with different polarization states. (**C**) Schematic of a representative demonstration. Line segments and ellipses represent local polarization profiles. (**D**) Two measured independent 3D vector field profiles. The top and bottom panels of each input state are profiles measured at different transverse planes, $z_1$ = 480 μm and $z_2$ = 960 μm, respectively.

## 2.4. Proposal for Arbitrary Linear Optical Transformation

With the proposed universal metasurfaces, one can construct a complete vector linear optical platform that can perform arbitrary vector linear operations on incident beams. It was previously shown that arbitrary scalar (i.e., polarization-insensitive) linear optical transformation can be constructed using only diffractive optical components and lenses[40-43]. Mathematically, this is equivalent to the proof of decomposability of an arbitrary matrix representing a linear operator $\mathcal{L}: X \to X'$ into diagonal matrices (*D*) and discrete Fourier transform matrices (*F*), where *X* is a scalar field vector space. We expand this result and prove that an arbitrary passive vector linear optical transformation (an arbitrary passive matrix for $\mathcal{L}: (X \otimes P) \to (X \otimes P)'$, where *P* is the polarization state vector space) can be constructed using only universal metasurfaces ($D^{(2)}$: block diagonal matrices with 2-by-2 sized blocks representing the Jones matrix at each spatial position) and conventional lenses ($F^{(2)} = F \otimes [1,\ 0;\ 0,\ 1]$) (Supplementary text S2).

As a specific example of such vectorial linear optical devices, we propose a probabilistic linear quantum optical CNOT gate composed of three universal metasurfaces and two lenses (Fig. 5A). The metasurface-based platform can be parallelized in two-dimensional arrays (Fig. 5B) and also enables manipulation of transverse modes, which is challenging with PICs. A transmission matrix of the CNOT gate in Ref. 35 represented in $X \otimes P$ basis, can be decomposed as follows:

$$\frac{1}{\sqrt{3}}\begin{pmatrix} 1 & 0 & 0 & 0 \\ 0 & -1 & 1 & 1 \\ 0 & 1 & 1 & 0 \\ 0 & 1 & 0 & 1 \end{pmatrix} = D_1^{(2)} F^{(2)} D_2^{(2)} F^{(2)} D_3^{(2)}\ , \tag{5}$$

and we design a metasurface-based CNOT gate using this factorization and assume ideal universal metasurfaces (Methods). The designed CNOT gate has additional controllability on orbital angular momentum (OAM) of photons[44] compare to that of the previous implementation in Ref. 35, and this enables rotation-tolerant qubit encoding[5].

We estimate the performance of the CNOT gate with Fourier optics analysis (Figs. 5C–F) (Methods). The calculated input-output field profiles (Fig. 5C) and the resultant transmission matrix (Fig. 5D) confirm the intended behavior of the device. The calculated transmission matrix is very similar to Eq. 5, although the overall magnitude decreases by a factor of 1.42, mainly due to the discrepancy between the mathematical discrete Fourier transform and the optical Fourier transform performed by conventional lenses (Methods). The calculated truth table also verifies proper CNOT gate functionality (Fig. 5E). The density matrix of the output state for the input state $(|0\rangle - |1\rangle)|1\rangle/\sqrt{2}$ is shown in Fig. 5F, and the fidelity with state $(|01\rangle - |10\rangle)/\sqrt{2}$ becomes close to 100%, confirming the ability of the proposed CNOT gate to create an entangled pair.

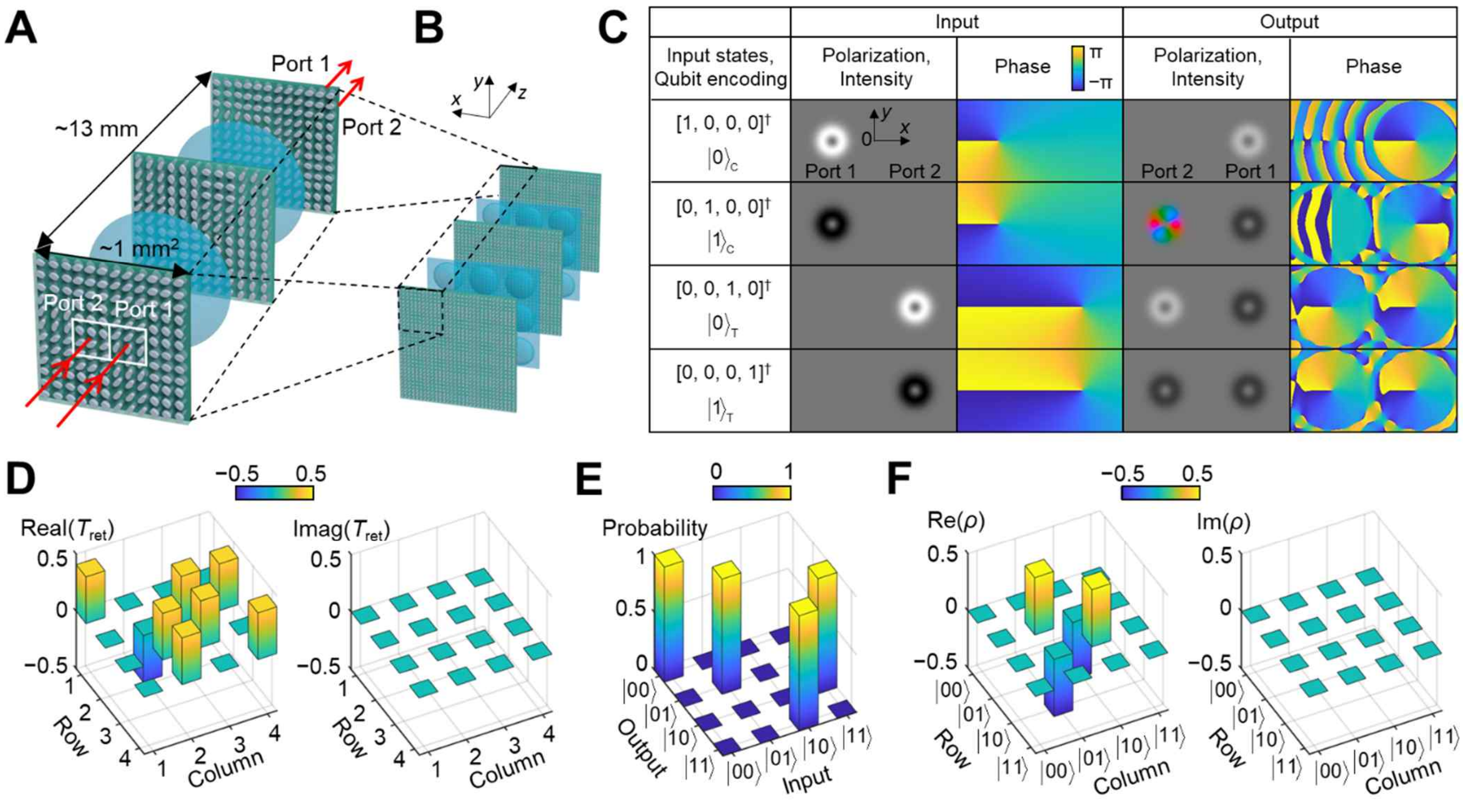

**Figure 5. Proposed metasurface-based probabilistic linear quantum optical CNOT gate.** (**A**) Schematic of proposed CNOT gate (not drawn to scale). Universal metasurfaces are placed at focal points of a 4f system. (**B**) Integration scheme using micro-lens arrays. (**C**) Polarization, intensity, and phase profiles of input and output states at corresponding ports, as described in (**A**). $|p\rangle_{\mathrm{C}}$ and $|q\rangle_{\mathrm{T}}$ ($p,\ q = 0,\ 1$) are the control qubit and target qubit, respectively. (**D**) Retrieved transmission matrix. (**E**) Retrieved truth table. $|pq\rangle$ refers to $|p\rangle_{\mathrm{C}} \otimes |q\rangle_{\mathrm{T}}$. (**F**) A retrieved density matrix for the input state of $(|0\rangle - |1\rangle)|1\rangle/\sqrt{2}$.

## 3. Conclusion

In this study, we achieved complete pixel-wise control of transmitted coherent light by devising a systematic way to realize the most general chiral and anisotropic metasurfaces that can implement arbitrary passive Jones matrices in a pixel-wise manner. First, we theoretically proved that a bilayer array of nanostructures is sufficient to cover the full set of passive Jones matrices. The extended degree of freedom of Jones matrices that universal metasurfaces provide makes possible unprecedented optical functionalities. We presented examples of new types of holographic imaging schemes, each of which breaks limitations of holographic images generated by conventional metasurfaces especially for the independent controllability of output polarizations, amplitudes, and phases for two different input polarization states. By fabricating bi-layer arrays of silicon posts, we demonstrated that all three aspects—polarization, amplitudes, and phases—of two different three-dimensional light field distributions are independently controllable as long as they are legitimate solutions of Maxwell's equations in a source-free half-space. In addition, we proposed a new optical platform composed of universal metasurfaces and lenses that can realize arbitrary passive vector linear optical devices, including an array of probabilistic linear quantum optical CNOT gates.

We note that the performance of devices based on universal metasurface can be further enhanced with optimization. We utilized a direct design method, where the proper structural parameters can be derived from analytic expressions based on the desired Jones matrix at each region of the metasurface. This method is conceptually clear and computationally very efficient. However, one can also utilize numerical optimizations considering near-field coupling between different unit structures and its effect on the transmitted amplitude and phase for each polarization. While doing so globally over the entire metasurface area is computationally very intensive, there are recent proposals toward fast optimization methods that are potentially applicable to large-area metasurfaces, such as gradient-based optimization[45-47] or machine learning-based approaches[47]. In addition, we would like to emphasize that the universal metasurfaces presented here provide full controllability for both input polarization channels for the first time, compared to existing metasurfaces giving full controllability only for one of the two possible input polarization channels. This could lead to a new way of controlling three-dimensional light fields in real time. Overall, we expect that the fully extended set of optical properties provided by the universal metasurfaces suggested here may find uses in many different fields.

# 4. Methods (4.1–4.10)

### 4.1. Fabrication of universal (including unitary) metasurfaces

Amorphous silicon (a-Si) thin film (thickness of 790 nm) is deposited on the fused silica substrate with plasma-enhanced chemical vapor deposition (PECVD) for the lower metasurface layer. In order to do aligned electron-beam (e-beam) lithography for a lower and an upper layer, an align-key is fabricated on the a-Si layer using additional e-beam lithography followed by metal deposition using e-beam evaporation (20 nm of Cr followed by 50 nm of Au) and lift-off. The lower layer of a cylinder array (square lattice of 450 nm period) is defined by aligned e-beam lithography using an align-key. Alumina hard mask (70 nm thickness) is realized with e-beam evaporation deposition and lift-off. Pseudo-Bosch dry etching process is utilized to realize the vertical side-wall slope of nano-pillar structures. The remaining alumina hard mask is utilized as a part of cylinders. Entire lower layer is encapsulated and planarized by 1.3 um thickness of SU8 polymer by spin coating and hardening. Additional a-Si layer thin film (thickness of 650 nm) is deposited with radio-frequency (RF) sputtering deposition for the upper metasurface layer. The upper metasurface layer is fabricated as a similar process for the lower layer.

### 4.2. Finite-difference time-domain (FDTD) simulation of unitary metasurface to optimize geometric parameters

A bilayer of ideal elliptic cylinder arrays (schematics in Fig. 1B) composed of realistic materials is assumed for a unitary metasurface in FDTD simulations. Throughout this study, commercial software from Ansys Lumerical Inc. is utilized for FDTD simulations. Based on the ellipsometry measurement, a complex refractive index of $n_{\text{c-Si}} + 3i\kappa_{\text{c-Si}}$ where $n_{\text{c-Si}}$ and $\kappa_{\text{c-Si}}$ are the real and imaginary index of crystalline silicon is assumed for PECVD a-Si, and complex refractive index of $3.7 + 0.05i$ is assumed for RF sputtered a-Si at 915 nm wavelength. A refractive index of 1.56 is assumed for the hardened SU8 polymer. All electromagnetic field data are obtained for 915 nm wavelength.

First, square lattice with 450 nm × 450 nm period, 790 nm thickness for layer A, 650 nm thickness for layer B are chosen for best performance based on the FDTD simulation results. Next, we find optimized pairs of $l_{AM}$ and $l_{Am}$ which realize every 36 combinations of $\phi_{AM}$ = {−9π/12, −5π/12, −π/12, 3π/12, 7π/12, 11π/12} and $\phi_{Am}$ = {−10π/12, −6π/12, −2π/12, 2π/12, 6π/12, 10π/12} for the 790 nm-thick for PECVD a-Si elliptic cylinders by sweeping fine grid of all $l_{AM}$ and $l_{Am}$ combinations in the FDTD simulations of SU8-encapsulated layer A. Similarly, we find an optimized pair of $l_{BM}$ and $l_{Bm}$ which realize $\phi_{BM} = 0$ and $\phi_{Bm} = \pi$ for the 650 nm-thick RF sputtered a-Si elliptic cylinders (global phase is omitted for both layer A and layer B).

### 4.3. Comparison between target optical properties and FDTD simulation results for unitary metasurfaces in Fig. 2A–B

For the FDTD simulation data in Fig. 2, both elliptic cylinders in layer A and layer B are assumed to be composed of PECVD a-Si in order to investigate achievable performance limit for the unitary metasurfaces (RF sputtered a-Si is more absorptive than PECVD a-Si). Based on the analytic solution in Supplementary text S1, we design 36 unitary metasurfaces (each composed of unit structures with an identical shape, i.e., spatially homogeneous) to have the same polarization conversion but to have all different accumulated phase pairs. Designed output polarization states are $|w\rangle$ and $|w_{\perp}\rangle$ under the incidence of right- and left-handed circular polarization states (RCP and LCP), respectively, where the spherical coordinate of $|w\rangle$ on the Poincaré sphere is $(2\psi,\ 2\chi) = (2\pi/3,\ -\pi/12)$.

### 4.4. Comparison between target Jones matrices and FDTD simulations results for universal metasurfaces in Fig. 2C–D

Similar to the FDTD simulations for Fig. 2A–B, both elliptic cylinders in layer A and layer B are assumed to be composed of PECVD a-Si. We utilized a 15-by-15 cluster of unitary metasurface unit structures (~7$\lambda$ in length) as one part of unit structure of universal metasurface to minimize near-field effects in this simulation. As stated in the main text, clusters size can be reduced without deterioration of performance by another optimization for shapes of elliptic cylinders to compensate for near-field effects, which is not in the scope of this study. Since we can realize an arbitrary Jones matrix whose maximum singular value is less than 0.85 with a-Si in the FDTD simulations as stated in the Supplementary text S1, we applied this constraint when we chose random and modified target Jones matrices.

### 4.5. Generation of two orthogonal Poincaré beam with independent phase profiles shown in Fig. 3B

Spatially varying polarization states of two Poincaré beams are designed based on Ref. 36 for RCP and LCP input. Phase profiles are designed such that one output beam to have divergence angle of 1.4° and deflection angle of 5° toward $+y$ direction, and another output beam to have divergence angle of 0.7° and deflection angle of 10° toward $+x$ direction.

The input Gaussian beams are focused at the metasurface where the beam waist is around 37.5 μm for both RCP and LCP incidence. The output beam profiles are measured at the far-field at a finite distance using a convex lens. Spatially varying polarization states of a Poincaré beams are quantitatively retrieved with standard full-Stokes polarimetry as described in Fig. S1.

### 4.6. Generation of two independent scalar (i.e., homogeneous polarization state over the entire image area) holographic images shown in Fig. 3C

For each target far-field image, we find spatially varying output phase profiles $\exp i\phi^{(n)}$ using Gerchberg-Saxton (GS) algorithm where $\phi^{(n)} \in \mathbb{R}$, $n = 1,\ 2$ for each image. We set spatially identical output polarization state under RCP input as $|w\rangle$ where its spherical coordinate on the Poincare sphere is $2\psi = \pi/6$ and $2\chi = \pi/12$. Then, we construct spatially varying unitary Jones matrices for two independent scalar holographic images as $U(\mathbf{r}) = \begin{pmatrix} |w\rangle & |w\rangle_{\perp} \end{pmatrix} \begin{pmatrix} \exp i\phi^{(1)}(\mathbf{r}) & 0 \\ 0 & \exp i\phi^{(2)}(\mathbf{r}) \end{pmatrix} \cdot \frac{1}{\sqrt{2}} \begin{pmatrix} 1 & -i \\ 1 & i \end{pmatrix}$, which generate scalar holographic images of rose and butterfly under RCP and LCP incidence, respectively.

The input Gaussian beams are focused at the metasurface where the beam waist is around 300 μm for both RCP and LCP incidence. The output scalar holographic images are measured at the far-field at a finite distance using a convex lens. Output polarization states are quantitatively retrieved with standard full-Stokes polarimetry as described in Fig. S1.

### 4.7. Generation of two independent vectorial holographic images shown in Fig. 3D

For each target far-field image, using the modified GS algorithm based on Ref. 37, we find spatially varying output polarization states and phases at the metasurface plane, $[E_x^{(n)};\ E_y^{(n)}\exp i\zeta^{(n)}]\exp i\phi^{(n)}$ where $E_x^{(n)}, E_y^{(n)}, \zeta^{(n)}, \phi^{(n)} \in \mathbb{R}$ , $n = 1,\ 2$ for each image, and $\sqrt{\left|E_x^{(n)}\right|^2 + \left|E_y^{(n)}\right|^2} = 1$ for all spatial positions on the metasurface. Then, we construct spatially varying Jones matrices for two independent holographic images as $J(\mathbf{r}) = \frac{1}{J_0}\begin{pmatrix} E_x^{(1)}(\mathbf{r}) & E_x^{(2)}(\mathbf{r}) \\ E_y^{(1)}(\mathbf{r})\exp i\zeta^{(1)}(\mathbf{r}) & E_y^{(2)}(\mathbf{r})\exp i\zeta^{(2)}(\mathbf{r}) \end{pmatrix}\begin{pmatrix} \exp i\phi^{(1)}(\mathbf{r}) & 0 \\ 0 & \exp i\phi^{(2)}(\mathbf{r}) \end{pmatrix}\frac{1}{\sqrt{2}}\begin{pmatrix} 1 & -i \\ 1 & i \end{pmatrix}$ where normalization factor $J_0$ is a maximum singular value of $J(\mathbf{r})$ over the entire $\mathbf{r}$. A universal metasurface realizing this $J(\mathbf{r})$ generates vectorial holographic images of rose and butterfly under RCP and LCP incidence, respectively.

We note that singular values of Jones matrices at various spatial positions on the metasurface may vary because an inner product of two polarization states is not necessarily the same for various spatial positions.

The input Gaussian beams are focused at the metasurface where the beam waist is around 300 μm for both RCP and LCP incidence. The output vectorial holographic images are measured at the far-field at a finite distance using a convex lens. Spatially varying polarization states are quantitatively retrieved with standard full-Stokes polarimetry as described in Fig. S1.

### 4.8. Generation of two independent three-dimensional (3D) vector field profiles shown in Fig. 4D

For each target 3D vector field profile, we find spatially varying output polarization states, amplitude, and phases by adding complex conjugation of electric field pattern radiation from all Gaussian beams with desired polarization states whose waists (2–5 μm) are at the desired intensity hotspots.

The input Gaussian beams are roughly focused at the metasurface, where the beam waist is around 1 mm for both RCP and LCP incidence. The output 3D vector field profiles are measured with an imaging system composed of an objective lens and a tube lens. Images at different transverse planes are obtained by varying the position of the metasurface along the optic axis of the imaging system. Spatially varying polarization states of 3D vector field profiles are quantitatively retrieved with standard full-Stokes polarimetry as described in Fig. S1.

## 4.9. Design of a metasurface-based probabilistic linear optical controlled-NOT (CNOT) gate and Fourier optic calculation for Fig. 5C

A transmission matrix of a CNOT gate in Ref. 35 represented in $X \otimes P$ basis can be factorized in the form of Eq. 5 using a nonlinear least-square fitting algorithm provided by commercial software as follow:

$$\begin{aligned}\frac{1}{\sqrt{3}}\begin{pmatrix}1 & 0 & 0 & 0\\ 0 & -1 & 1 & 1\\ 0 & 1 & 1 & 0\\ 0 & 1 & 0 & 1\end{pmatrix} &= D_1^{(2)}F^{(2)}D_2^{(2)}F^{(2)}D_3^{(2)}\\ &= \begin{pmatrix}a_1 & 0 & 0 & 0\\ 0 & a_2 & 0 & 0\\ 0 & 0 & a_3 & a_4\\ 0 & 0 & -a_3 & a_4\end{pmatrix}F^{(2)}\begin{pmatrix}a_5 & 0 & 0 & 0\\ 0 & a_6 & 0 & 0\\ 0 & 0 & a_5 & 0\\ 0 & 0 & 0 & a_7\end{pmatrix}F^{(2)}\begin{pmatrix}a_1 & 0 & 0 & 0\\ 0 & a_2 & 0 & 0\\ 0 & 0 & a_3 & -a_3\\ 0 & 0 & a_4 & a_4\end{pmatrix}\end{aligned} \qquad (6)$$

where $a_1$ to $a_7$ are 0.8095+0.5405$i$, 0.6208+0.7839$i$, −0.5724−0.3822$i$, 0.5543−0.4390$i$, 0.2336−0.5628$i$, 0.9271+0.3749$i$, and −0.6625+0.7491$i$, respectively. Singular values of all 2-by-2 blocks (Jones matrices) are equal or less than 1, which means the block diagonal matrices can be realized with universal metasurfaces with arbitrary passive Jones matrices. However, the retrieved transmission matrix of a designed optical system based on the Eq. 6 using Fourier optic calculation (described at the end of this section) becomes $\begin{pmatrix} 0.5733 & 0 & 0 & 0 \\ 0 & -0.5732 & 0.3653 & 0.3653 \\ 0 & 0.3653 & 0.5733 & 0 \\ 0 & 0.3653 & 0 & 0.5733 \end{pmatrix}$ (imaginary parts of all elements are smaller than 1/1000), which is significantly different from the target transmission matrix. This difference is due to the discrepancy between the mathematical discrete Fourier transform with $F^{(2)}$ and the optical Fourier transform with a lens. While the mathematical discrete Fourier transform with $F^{(2)}$ inherently assumes transversal periodic copies of input states in the spatial domain, there is no signal except around the optic axis for an optical system. With trial and error scheme, we consider the following factorization,

$$\begin{aligned} \frac{1}{\sqrt{3}}\begin{pmatrix} 0.7099 & 0 & 0 & 0 \\ 0 & -0.7099 & 1.115 & 1.115 \\ 0 & 1.115 & 0.7099 & 0 \\ 0 & 1.115 & 0 & 0.7099 \end{pmatrix} &= D_1^{(2)\prime} F^{(2)} D_2^{(2)\prime} F^{(2)} D_3^{(2)\prime} \\ &= \begin{pmatrix} b_1 & 0 & 0 & 0 \\ 0 & b_2 & 0 & 0 \\ 0 & 0 & b_3 & b_4 \\ 0 & 0 & -b_3 & b_4 \end{pmatrix} F^{(2)} \begin{pmatrix} b_5 & 0 & 0 & 0 \\ 0 & b_6 & 0 & 0 \\ 0 & 0 & b_5 & 0 \\ 0 & 0 & 0 & b_7 \end{pmatrix} F^{(2)} \begin{pmatrix} b_1 & 0 & 0 & 0 \\ 0 & b_2 & 0 & 0 \\ 0 & 0 & b_3 & -b_3 \\ 0 & 0 & b_4 & b_4 \end{pmatrix} \end{aligned} \tag{7}$$

where $b_1$ to $b_7$ are $-0.7441+0.5429i$, $-0.9906+0.1340i$, $-0.5262+0.3839i$, $-0.0948-0.7004i$, $0.1475+0.4600i$, $-0.1534-0.9874i$, $-0.6375+0.7694i$, respectively. Singular values of all 2-by-2 blocks (Jones matrices) are less than 1, again, which means the block diagonal matrices can be realized with universal metasurfaces with arbitrary passive Jones matrices. The retrieved transmission matrix of a designed optical system based on the Eq. 7 using Fourier optic calculation becomes $\frac{0.7048}{\sqrt{3}}\begin{pmatrix} 1 & 0 & 0 & 0 \\ 0 & -1 & 1 & 1 \\ 0 & 1 & 1 & 0 \\ 0 & 1 & 0 & 1 \end{pmatrix}$ (where imaginary parts of all elements are smaller than 1/1000), which is the desired value, although the overall magnitude is decreased by a factor of 1.42.

In order to achieve rotation-tolerant qubit encoding, we consider the $X \otimes \{|R\rangle \otimes |l\rangle,\ |L\rangle \otimes |r\rangle\}$ basis rather than the $X \otimes P$ basis where $|l\rangle$ and $|r\rangle$ are Laguerre-gaussian modes of zero

radial index with −1 and +1 azimuthal index, respectively. By merging $q$-plates (Ref. 44) to both endmost universal metasurfaces, one channel of the proposed CNOT gate array does the same role in Ref. 35 for the $X \otimes \{|R\rangle \otimes |l\rangle,\ |L\rangle \otimes |r\rangle\}$ basis. We note that this qubit encoding does not increase system complexity because $q$-plate merged universal metasurfaces, $D_1^{(2)\prime\prime} = T_q D_1^{(2)\prime}$ and $D_3^{(2)\prime\prime} = D_3^{(2)\prime} T_q^{-1}$, are still single universal metasurfaces where $T_q$ is a transmission matrix of a $q$-plate. The $T_q^{-1}$ is designed to make the orbital angular momentums of all states within the CNOT gate to be zero. The angular momentums are restored at the exit of the CNOT gate because of $T_q$.

We design ideal universal metasurfaces, i.e., spatially varying arbitrary passive Jones matrices, and an overall optical system including lenses considering target maximum numerical aperture. The maximum numerical aperture is set to be a low enough value, 0.08, in order to allow a larger universal meta-atom size that is easier to optimize and experimentally realize in future works. Accordingly, we assume target beam waist, a single port size (Fig. 5A) and lens focal length to be 9 μm, 54 μm × 54 μm, and 3.2 mm, respectively, for the working wavelength of 915 nm. This gives device length along optic axis direction around 13 mm as in Fig. 5A. Designed metasurfaces are represented in Fig. S4.

We calculate the output field profile for various input states represented in Fig. 5C using Fourier optics considering polarization states, i.e., Jones vectors. For a given input field profile (spatially varying Jones vectors), we first calculate the field profiles right before the first lens by multiplying designed Jones matrices of the frontend universal metasurface ($D_3^{(2)}$) followed by free-space propagation to the first lens. After applying the phase profile of the first lens, consider another free-space propagation to the next metasurface. A similar process can be done until we obtain the field profile after the last metasurface. We consider large enough zero-padding for the field profiles, which determines enough transversal device area to avoid crosstalk in case of integration (Fig. 5B) around 1 mm$^2$ as in Fig. 5A.

### 4.10. Retrieval of the transmission matrix, truth table, and the density matrix of a highly entangled state for the proposed CNOT gate (Fig. 5D–F).

In order to retrieve the transmission matrix in the $X \otimes \{|R\rangle \otimes |l\rangle,\ |L\rangle \otimes |r\rangle\}$ basis, we calculate each element of the transmission matrix with mode overlap integral, $t_{ij} = \dfrac{\left|\iint_S E_i^*(\mathbf{r}) E_{\text{out}}^{(j)}(\mathbf{r}) dxdy\right|^2}{\iint_S |E_i(\mathbf{r})|^2 dxdy \iint_S |E_j(\mathbf{r})|^2 dxdy}$, where $E_i$ and $E_j$ are basis mode field profiles and $E_{\text{out}}^{(j)}$ is an output field profile under $E_j$ incidence based on Fourier optic calculations, which are shown in Fig. 5C. Output port number is flipped in the *x*-direction, as shown in Fig. 5A, due to the intrinsic image inversion property of the 4f system. This result is shown in Fig. 5D.

As long as we know transmission matrix and logical qubit basis is given, truth table and output qubit states (and density matrices) can be calculated, as an example in Ref. 2. We set rotation-tolerant logical qubit encoding such that single-photon states in port 1 and port 2 as a control and a target qubit, respectively: $|R\rangle \otimes |l\rangle$, and $|L\rangle \otimes |r\rangle$ state as $|0\rangle$ and $|1\rangle$, respectively, as shown in Fig. 5C.

First, the truth table is retrieved based on the transmission matrix. The relationship between creation operators for the input and output state becomes,

$$
\begin{aligned}
a_{\text{in}}^\dagger &\to t_{11} a_{\text{out}}^\dagger + t_{21} b_{\text{out}}^\dagger + t_{31} c_{\text{out}}^\dagger + t_{41} d_{\text{out}}^\dagger \\
b_{\text{in}}^\dagger &\to t_{12} a_{\text{out}}^\dagger + t_{22} b_{\text{out}}^\dagger + t_{32} c_{\text{out}}^\dagger + t_{42} d_{\text{out}}^\dagger \\
c_{\text{in}}^\dagger &\to t_{13} a_{\text{out}}^\dagger + t_{23} b_{\text{out}}^\dagger + t_{33} c_{\text{out}}^\dagger + t_{43} d_{\text{out}}^\dagger \\
d_{\text{in}}^\dagger &\to t_{14} a_{\text{out}}^\dagger + t_{24} b_{\text{out}}^\dagger + t_{34} c_{\text{out}}^\dagger + t_{44} d_{\text{out}}^\dagger
\end{aligned}
\qquad (8)
$$

where $a_{\text{in(out)}}^\dagger$, $b_{\text{in(out)}}^\dagger$, $c_{\text{in(out)}}^\dagger$, and $d_{\text{in(out)}}^\dagger$ are creation operators for single-photon input (output) states $(1,\ 0,\ 0,\ 0)^\dagger$, $(0,\ 1,\ 0,\ 0)^\dagger$, $(0,\ 0,\ 1,\ 0)^\dagger$, and $(0,\ 0,\ 0,\ 1)^\dagger$ in $X \otimes \{|R\rangle \otimes |l\rangle,\ |L\rangle \otimes |r\rangle\}$ basis, respectively. Output qubit state under input qubit state of $|pq\rangle$ $(= |p\rangle_{\text{C}} \otimes |q\rangle_{\text{T}}$, *p*, *q* = 0, 1) can be calculated with Eq. 8. For example, an output qubit state under the input state of $|00\rangle_{\text{in}}$ becomes $(t_{11} a_{\text{out}}^\dagger + t_{21} b_{\text{out}}^\dagger + t_{31} c_{\text{out}}^\dagger + t_{41} d_{\text{out}}^\dagger)(t_{13} a_{\text{out}}^\dagger + t_{23} b_{\text{out}}^\dagger + t_{33} c_{\text{out}}^\dagger + t_{43} d_{\text{out}}^\dagger)$. Among 16 terms, eight terms representing single-photon outcome per each port (control and target) are considered successful operations, which becomes $\left(t_{11}t_{33}+t_{31}t_{13}\right)\left|00\right\rangle_{\text{out}}+\left(t_{11}t_{43}+t_{41}t_{13}\right)\left|01\right\rangle_{\text{out}}+\left(t_{21}t_{33}+t_{31}t_{23}\right)\left|10\right\rangle_{\text{out}}+\left(t_{21}t_{43}+t_{41}t_{23}\right)\left|11\right\rangle_{\text{out}}$. Therefore, the complex amplitude of a truth table becomes,

$$C_{\text{TT}}=\begin{pmatrix} t_{11}t_{33}+t_{31}t_{13} & t_{11}t_{34}+t_{31}t_{14} & t_{12}t_{33}+t_{32}t_{13} & t_{12}t_{34}+t_{32}t_{14} \\ t_{11}t_{43}+t_{41}t_{13} & t_{11}t_{44}+t_{41}t_{14} & t_{12}t_{43}+t_{42}t_{13} & t_{12}t_{44}+t_{42}t_{14} \\ t_{21}t_{33}+t_{31}t_{23} & t_{21}t_{34}+t_{31}t_{24} & t_{22}t_{33}+t_{32}t_{23} & t_{22}t_{34}+t_{32}t_{24} \\ t_{21}t_{43}+t_{41}t_{23} & t_{21}t_{44}+t_{41}t_{24} & t_{22}t_{43}+t_{42}t_{23} & t_{22}t_{44}+t_{42}t_{24} \end{pmatrix}. \quad (7)$$

The probability of a truth table can be calculated by the element-wise square of each element's magnitude. One can normalize each column with its magnitude in order to obtain conditional probability upon success. This conditional probability of truth table is represented in Fig. 5E. For the proposed CNOT gate, the probability of success (normalization value of each column) is around $\frac{1}{9}\times\frac{1}{4.05}$ (same for all columns), where $\frac{1}{9}$ is the fundamental probability limit.

Since we know complex amplitudes of input-output qubit states transformations (i.e., complex amplitudes of a truth table), we can directly calculate output states $\left|\psi\right\rangle$ as well as density matrix $\left|\psi\right\rangle\left\langle\psi\right|$ using this relation. As an example in Fig. 5F, the output state under the input state $\left(\left|0\right\rangle-\left|1\right\rangle\right)\left|1\right\rangle/\sqrt{2}$ can be calculated as $\left|\psi\right\rangle=C_{\text{TT}}\left(\left|01\right\rangle-\left|11\right\rangle\right)/\sqrt{2}$ and can be normalized with its magnitude. The result density matrix $\left|\psi\right\rangle\left\langle\psi\right|$ is shown in Fig. 5F, and fidelity with $\left(\left|01\right\rangle-\left|10\right\rangle\right)/\sqrt{2}$ can be calculated as $\frac{1}{2}\left(\left\langle 01\right|-\left\langle 10\right|\right)\left|\psi\right\rangle\left\langle\psi\right|\left(\left|01\right\rangle-\left|10\right\rangle\right)$, which gives almost 100% as noted in the main text.

## 5. References

1. Simon, R. & Mukunda, N. Minimal three-component SU(2) gadget for polarization optics. *Phys. Lett. A* **143**, 165-169 (1990).

2. Kok, P. et al. Linear optical quantum computing with photonic qubits. *Rev. Mod. Phys.* **79**, 135-174 (2007).

3. Rubin, N. A. et al. Matrix Fourier optics enables a compact full-Stokes polarization camera. *Science* **365**, eaax1839 (2019).

4. Kwon, H., Arbabi, E., Kamali, S. M., Faraji-Dana, M. & Faraon, A. Single-shot quantitative phase gradient microscopy using a system of multifunctional metasurfaces. *Nat. Photonics* **14**, 109-114 (2020).

5. D'Ambrosio, V. et al. Complete experimental toolbox for alignment-free quantum communication. *Nat. Commun.* **3**, 961 (2012).

6. Bozinovic, N. et al. Terabit-Scale Orbital Angular Momentum Mode Division Multiplexing in Fibers. *Science* **340**, 1545-1548 (2013).

7. Devlin, R. C., Ambrosio, A., Rubin, N. A., Mueller, J. P. B. & Capasso, F. Arbitrary spin-to–orbital angular momentum conversion of light. *Science* **358**, 896-901 (2017).

8. Stav, T. et al. Quantum entanglement of the spin and orbital angular momentum of photons using metamaterials. *Science* **361**, 1101-1104 (2018).

9. Menzel, C., Rockstuhl, C. & Lederer, F. Advanced Jones calculus for the classification of periodic metamaterials. *Phys. Rev. A* **82**, 053811 (2010).

10. Rubin, N. A., Zaidi, A., Dorrah, A. H., Shi, Z. & Capasso, F. Jones matrix holography with metasurfaces. *Sci. Adv.* **7**, eabg7488 (2021).

11. Plum, E., Fedotov, V. A. & Zheludev, N. I. Planar metamaterial with transmission and reflection that depend on the direction of incidence. *Appl. Phys. Lett.* **94**, 131901 (2009).

12. Yu, N. et al. Light Propagation with Phase Discontinuities: Generalized Laws of Reflection and Refraction. *Science* **334**, 333-337 (2011).

13. Huang, L. et al. Three-dimensional optical holography using a plasmonic metasurface. *Nat. Commun.* **4**, 2808 (2013).

14. Arbabi, A., Horie, Y., Bagheri, M. & Faraon, A. Dielectric metasurfaces for complete control of phase and polarization with subwavelength spatial resolution and high transmission. *Nat. Nanotechnol.* **10**, 937-943 (2015).

15. Balthasar Mueller, J. P., Rubin, N. A., Devlin, R. C., Groever, B. & Capasso, F. Metasurface Polarization Optics: Independent Phase Control of Arbitrary Orthogonal States of Polarization. *Phys. Rev. Lett.* **118**, 113901 (2017).

16. Deng, Z.-L. et al. Diatomic Metasurface for Vectorial Holography. *Nano Lett.* **18**, 2885-2892 (2018).

17. Song, Q. et al. Broadband decoupling of intensity and polarization with vectorial Fourier metasurfaces. *Nat. Commun.* **12**, 3631 (2021).

18. Lung, S. et al. Complex-Birefringent Dielectric Metasurfaces for Arbitrary Polarization-Pair Transformations. *ACS Photonics* **7**, 3015-3022 (2020).

19. Dorrah, A. H., Rubin, N. A., Zaidi, A., Tamagnone, M. & Capasso, F. Metasurface optics for on-demand polarization transformations along the optical path. *Nat. Photonics* **15**, 287-296 (2021).

20. Gansel, J. K. et al. Gold Helix Photonic Metamaterial as Broadband Circular Polarizer. *Science* **325**, 1513-1515 (2009).

21. Decker, M. et al. Strong optical activity from twisted-cross photonic metamaterials. *Opt. Express* **34**, 2501-2503 (2009).

22. Tanaka, K. et al. Chiral Bilayer All-Dielectric Metasurfaces. *ACS Nano* **14**, 15926-15935 (2020).

23. Liu, N., Liu, H., Zhu, S. & Giessen, H. Stereometamaterials. *Nat. Photonics* **3**, 157-162 (2009).

24. Menzel, C. et al. Asymmetric Transmission of Linearly Polarized Light at Optical Metamaterials. *Phys. Rev. Lett.* **104**, 253902 (2010).

25. Zhao, Y., Belkin, M. A. & Alù, A. Twisted optical metamaterials for planarized ultrathin broadband circular polarizers. *Nat. Commun.* **3**, 870 (2012).

26. Huang, Y.-W. et al. Versatile total angular momentum generation using cascaded J-plates. *Opt. Express* **27**, 7469-7484 (2019).

27. Plum, E., Fedotov, V. A. & Zheludev, N. I. Optical activity in extrinsically chiral metamaterial. *Appl. Phys. Lett.* **93**, 191911 (2008).

28. Shi, Z. et al. Continuous angle-tunable birefringence with freeform metasurfaces for arbitrary polarization conversion. *Sci. Adv.* **6**, eaba3367 (2020).

29. Cerjan, A. & Fan, S. Achieving Arbitrary Control over Pairs of Polarization States Using Complex Birefringent Metamaterials. *Phys. Rev. Lett.* **118**, 253902 (2017).

30. Lee, G.-Y. et al. Complete amplitude and phase control of light using broadband holographic metasurfaces. *Nanoscale* **10**, 4237-4245 (2018).

31. Hsueh, C. K. & Sawchuk, A. A. Computer-generated double-phase holograms. *Appl. Opt.* **17**, 3874-3883 (1978).

32. Ngcobo, S., Litvin, I., Burger, L. & Forbes, A. A digital laser for on-demand laser modes. *Nat. Commun.* **4**, 2289 (2013).

33. Mendoza-Yero, O., Mínguez-Vega, G. & Lancis, J. Encoding complex fields by using a phase-only optical element. *Opt. Lett.* **39**, 1740-1743 (2014).

34. Führ, H. & Rzeszotnik, Z. A note on factoring unitary matrices. *Linear Algebra Appl.* **547**, 32-44 (2018).

35. O'Brien, J. L., Pryde, G. J., White, A. G., Ralph, T. C. & Branning, D. Demonstration of an all-optical quantum controlled-NOT gate. *Nature* **426**, 264-267 (2003).

36. Beckley, A. M., Brown, T. G. & Alonso, M. A. Full Poincaré beams. *Opt. Express* **18**, 10777-10785 (2010).

37. Arbabi, E., Kamali, S. M., Arbabi, A. & Faraon, A. Vectorial Holograms with a Dielectric Metasurface: Ultimate Polarization Pattern Generation. *ACS Photonics* **6**, 2712-2718 (2019).

38. Ren, H., Shao, W., Li, Y., Salim, F. & Gu, M. Three-dimensional vectorial holography based on machine learning inverse design. *Sci. Adv.* **6**, eaaz4261 (2020).

39. Overvig, A. C. et al. Dielectric metasurfaces for complete and independent control of the optical amplitude and phase. *Light Sci. Appl.* **8**, 92 (2019).

40. Müller-Quade, J., Aagedal, H., Beth, T. & Schmid, M. Algorithmic design of diffractive optical systems for information processing. *Physica D* **120**, 196-205 (1998).

41. Schmid, M., Steinwandt, R., Müller-Quade, J., Rötteler, M. & Beth, T. Decomposing a matrix into circulant and diagonal factors. *Linear Algebra Appl.* **306**, 131-143 (2000).

42. Huhtanen, M. & Perämäki, A. Factoring Matrices into the Product of Circulant and Diagonal Matrices. *J. Fourier Anal. Appl.* **21**, 1018-1033 (2015).

43. Kulce, O., Mengu, D., Rivenson, Y. & Ozcan, A. All-optical synthesis of an arbitrary linear transformation using diffractive surfaces. *Light Sci. Appl.* **10**, 196 (2021).

44. Marrucci, L., Manzo, C. & Paparo, D. Optical Spin-to-Orbital Angular Momentum Conversion in Inhomogeneous Anisotropic Media. *Phys. Rev. Lett.* **96**, 163905 (2006).

45. Mansouree, M., McClung, A., Samudrala, S. & Arbabi, A. Large-Scale Parameterized Metasurface Design Using Adjoint Optimization. *ACS Photonics* **8**, 455-463 (2021).

46. Zhou, M. et al. Inverse Design of Metasurfaces Based on Coupled-Mode Theory and Adjoint Optimization. *ACS Photonics* **8**, 2265-2273 (2021).

47. Jiang, J., Fan, J. A. Global Optimization of Dielectric Metasurfaces Using a Physics-Driven Neural Network. *Nano Lett.* **19**, 5366-5372 (2019).

## Supplementary Text

**S1. Find an analytic relationship between geometric parameters and output polarization states with phases.**

We first optimized $l_{\mathrm{AM}}$, $l_{\mathrm{Am}}$, $l_{\mathrm{BM}}$, and $l_{\mathrm{Bm}}$ for the various target $\phi_{\mathrm{AM}}$, $\phi_{\mathrm{Am}}$, $\phi_{\mathrm{BM}}$, and $\phi_{\mathrm{Bm}}$ where $\phi_{i\mathrm{M}}$ and $\phi_{i\mathrm{m}}$ are phase advance for input polarization state along the major and minor axis, respectively, for layer *i* (A or B). Then, we treat $\phi_{\mathrm{AM}}$, $\phi_{\mathrm{Am}}$, $\phi_{\mathrm{BM}}$, and $\phi_{\mathrm{Bm}}$ as geometric parameters. Based on Eq. 2 in the main text, an arbitrary unitary Jones matrix can be represented as

$$P=\begin{pmatrix}\cos\theta'_{\mathrm{B}} & \sin\theta'_{\mathrm{B}}\\ -\sin\theta'_{\mathrm{B}} & \cos\theta'_{\mathrm{B}}\end{pmatrix}\begin{pmatrix}e^{i\phi_{\mathrm{AM}}} & 0\\ 0 & e^{i(\phi_{\mathrm{Am}}+\pi)}\end{pmatrix}\begin{pmatrix}\cos\theta_{\mathrm{A}} & -\sin\theta_{\mathrm{A}}\\ \sin\theta_{\mathrm{A}} & \cos\theta_{\mathrm{A}}\end{pmatrix} \quad \text{(S1)}$$

where $\theta'_{\mathrm{B}}=2\theta_{\mathrm{B}}-\theta_{\mathrm{A}}$. Meanwhile, an arbitrary unitary Jones matrix in terms of the input-output polarizations pairs and corresponding phase advances becomes

$$Q=\begin{pmatrix}\cos\psi & -\sin\psi\\ \sin\psi & \cos\psi\end{pmatrix}\begin{pmatrix}\cos\chi & \sin\chi\\ i\sin\chi & -i\cos\chi\end{pmatrix}\begin{pmatrix}e^{-i\phi'_1} & 0\\ 0 & e^{-i\phi'_2}\end{pmatrix}\begin{pmatrix}1/\sqrt{2} & -i/\sqrt{2}\\ 1/\sqrt{2} & i/\sqrt{2}\end{pmatrix} \quad \text{(S2)}$$

as described in the main text. Since solving $P=Q$ is not straightforward, we consider $P^{\mathrm{T}}P=Q^{\mathrm{T}}Q$ and $PP^{\mathrm{T}}=QQ^{\mathrm{T}}$ instead. Equation S1 and Eq. S2 give,

$$
\begin{cases}
P^{\mathrm{T}}P = e^{i(\phi_{\mathrm{AM}}+\phi_{\mathrm{Am}})}\begin{pmatrix} \cos(\phi_{\mathrm{AM}}-\phi_{\mathrm{Am}})+i\cos 2\theta_{\mathrm{A}}\sin(\phi_{\mathrm{AM}}-\phi_{\mathrm{Am}}) & i\sin 2\theta_{\mathrm{A}}\sin(\phi_{\mathrm{AM}}-\phi_{\mathrm{Am}}) \\ i\cos 2\theta_{\mathrm{A}}\sin(\phi_{\mathrm{AM}}-\phi_{\mathrm{Am}}) & \cos(\phi_{\mathrm{AM}}-\phi_{\mathrm{Am}})-i\cos 2\theta_{\mathrm{A}}\sin(\phi_{\mathrm{AM}}-\phi_{\mathrm{Am}}) \end{pmatrix}, \\
PP^{\mathrm{T}} = e^{i(\phi_{\mathrm{AM}}+\phi_{\mathrm{Am}})}\begin{pmatrix} \cos(\phi_{\mathrm{AM}}-\phi_{\mathrm{Am}})+i\cos 2\theta'_{\mathrm{B}}\sin(\phi_{\mathrm{AM}}-\phi_{\mathrm{Am}}) & i\sin 2\theta'_{\mathrm{B}}\sin(\phi_{\mathrm{AM}}-\phi_{\mathrm{Am}}) \\ i\cos 2\theta'_{\mathrm{B}}\sin(\phi_{\mathrm{AM}}-\phi_{\mathrm{Am}}) & \cos(\phi_{\mathrm{AM}}-\phi_{\mathrm{Am}})-i\cos 2\theta'_{\mathrm{B}}\sin(\phi_{\mathrm{AM}}-\phi_{\mathrm{Am}}) \end{pmatrix}, \\
Q^{\mathrm{T}}Q = e^{i(\phi'_1+\phi'_2)}\begin{pmatrix} \sin 2\chi + i\cos 2\chi\sin(\phi'_1-\phi'_2) & -i\cos 2\chi\cos(\phi'_1-\phi'_2) \\ -i\cos 2\chi\cos(\phi'_1-\phi'_2) & \sin 2\chi - i\cos 2\chi\sin(\phi'_1-\phi'_2) \end{pmatrix}, \\
QQ^{\mathrm{T}} = e^{i(\phi'_1+\phi'_2)}\begin{pmatrix} \sin 2\chi + i\sin 2\psi\cos 2\chi & -i\cos 2\psi\cos 2\chi \\ -i\cos 2\psi\cos 2\chi & \sin 2\chi - i\sin 2\psi\cos 2\chi \end{pmatrix}.
\end{cases}
\tag{S3}
$$

By making $P^{\mathrm{T}}P = Q^{\mathrm{T}}Q$ and $PP^{\mathrm{T}} = QQ^{\mathrm{T}}$ , we can find a trial solution if we assume $\phi_{\mathrm{AM}} + \phi_{\mathrm{Am}} = \phi'_1 + \phi'_2 + 3\pi$ ,

$$
\begin{cases}
\theta_{\mathrm{A}} = \dfrac{\phi'_1-\phi'_2}{2} + \dfrac{\pi}{4}, \\
\theta_{\mathrm{B}} = \dfrac{\psi}{2} + \dfrac{\phi'_1-\phi'_2}{4} + \dfrac{\pi}{4}, \\
\phi_{\mathrm{AM}} = \dfrac{\phi'_1+\phi'_2}{2} + \chi + \dfrac{7\pi}{4}, \\
\phi_{\mathrm{Am}} = \dfrac{\phi'_1+\phi'_2}{2} - \chi + \dfrac{5\pi}{4}.
\end{cases}
\tag{S4}
$$

Then, it can be verified that this trial solution satisfies $P = Q$, and this trial solution is a correct solution.

Equation S4 shows that there is a linear relationship between geometric parameters and output polarization state with phases under RCP and LCP incidence. Therefore, output polarization states on the Poincaré sphere and phases pairs are regularly spaced for a regular array of geometric parameters. By choosing a set of geometric parameters such as

$$\begin{cases} \theta_{\mathrm{A}} = \left\{ \frac{\pi}{12}, \frac{3\pi}{12}, \frac{5\pi}{12}, \frac{7\pi}{12}, \frac{9\pi}{12}, \frac{11\pi}{12} \right\}, \\ \theta_{\mathrm{B}} = \left\{ 0, \frac{\pi}{24}, \frac{2\pi}{24}, \cdots, \frac{22\pi}{24}, \frac{23\pi}{24} \right\}, \\ \phi_{\mathrm{AM}} = \left\{ -\frac{9\pi}{12}, -\frac{5\pi}{12}, -\frac{\pi}{12}, \frac{3\pi}{12}, \frac{7\pi}{12}, \frac{11\pi}{12} \right\}, \\ \phi_{\mathrm{Am}} = \left\{ -\frac{10\pi}{12}, -\frac{6\pi}{12}, -\frac{2\pi}{12}, \frac{2\pi}{12}, \frac{6\pi}{12}, \frac{10\pi}{12} \right\}, \end{cases} \tag{S5}$$

we can approximate arbitrary target unitary Jones matrix such as $U_{\mathrm{t}} \approx [U_{\mathrm{t}}]$ where $[U_{\mathrm{t}}]$ is a realizable unitary Jones matrix with the above geometric parameters. With a finer grid of geometric parameters, one can approximate an arbitrary unitary Jones matrix more precisely.

Since arbitrary passive Jones matrix can be represented as a sum of two unitary Jones matrix as described in the main text, a universal meta-atom of a target arbitrary passive Jones matrices, $J_{\mathrm{t}} = U_{\mathrm{t1}} + U_{\mathrm{t2}}$, can be designed based on $[U_1]$ and $[U_2]$. With rigorous consideration of transmission amplitudes, one can achieve $[U_{\mathrm{t}}] \approx \alpha U_{\mathrm{t}}$, where a loss factor $\alpha$ is around 0.85 for the FDTD simulations, if we assume lossy materials such as a-Si for realistic universal metasurfaces. The loss factor is almost constant over various universal meta-atoms. Therefore, we considered $\frac{1}{\alpha} J_{\mathrm{t}} = U'_{\mathrm{t1}} + U'_{\mathrm{t2}}$ instead in order to realize $J_{\mathrm{t}}$, whose maximum singular value is smaller than 0.85. Then, $J_{\mathrm{t}}$ can be approximated as $[U'_{\mathrm{t1}}] + [U'_{\mathrm{t2}}] \approx \alpha (U'_{\mathrm{t1}} + U'_{\mathrm{t2}}) = \alpha \cdot \frac{1}{\alpha} J_{\mathrm{t}} = J_{\mathrm{t}}$ with a realistic universal meta-atom.

**S2. Expansion of the result of Ref. 40 in the main text to a polarization-sensitive version of arbitrary linear devices.**

By following Ref. 40 in the main text, we first prove that any arbitrary $2^{n+1}$-by-$2^{n+1}$ (for integer $n$) matrix can be decomposed into a number of block diagonal matrices and block circulant matrices of the same size using the induction hypothesis. In this prove, block sizes are all 2-by-2.

Let us consider $\begin{pmatrix} I & I \\ I & -I \end{pmatrix}$, $\begin{pmatrix} C^{(2)} & 0 \\ 0 & I \end{pmatrix}$, and $\begin{pmatrix} I & 0 \\ 0 & C^{(2)} \end{pmatrix}$, where $I$ and $C^{(2)}$ are a $2^n$-by-$2^n$ identity matrix and $2^n$-by-$2^n$ block circulant matrix. By considering the following equalities:

$$\begin{cases} \begin{pmatrix} I & I \\ I & -I \end{pmatrix} = \begin{pmatrix} I & 0 \\ 0 & -i\cdot I \end{pmatrix}\begin{pmatrix} I & i\cdot I \\ i\cdot I & I \end{pmatrix}\begin{pmatrix} I & 0 \\ 0 & -i\cdot I \end{pmatrix} \\ \begin{pmatrix} C^{(2)} & 0 \\ 0 & I \end{pmatrix} = \frac{1}{4}\begin{pmatrix} I & I \\ I & -I \end{pmatrix}\begin{pmatrix} C^{(2)}+I & C^{(2)}-I \\ C^{(2)}-I & C^{(2)}+I \end{pmatrix}\begin{pmatrix} I & I \\ I & -I \end{pmatrix} \\ \begin{pmatrix} I & 0 \\ 0 & C^{(2)} \end{pmatrix} = \frac{1}{4}\begin{pmatrix} I & I \\ I & -I \end{pmatrix}\begin{pmatrix} I+C^{(2)} & I-C^{(2)} \\ I-C^{(2)} & I+C^{(2)} \end{pmatrix}\begin{pmatrix} I & I \\ I & -I \end{pmatrix} \end{cases}, \quad \text{(S6)}$$

we now know that $\begin{pmatrix} I & I \\ I & -I \end{pmatrix}$, $\begin{pmatrix} C^{(2)} & 0 \\ 0 & I \end{pmatrix}$, and $\begin{pmatrix} I & 0 \\ 0 & C^{(2)} \end{pmatrix}$ can be decomposed into block diagonal matrices and block circulant matrices. Therefore, $\begin{pmatrix} C^{(2)} & 0 \\ 0 & C'^{(2)} \end{pmatrix}$ also can be decomposed into block diagonal matrices and block circulant matrices, where $C'^{(2)}$ is another block circulant matrix. The induction hypothesis assumes that $\begin{pmatrix} A & 0 \\ 0 & B \end{pmatrix}$ can be decomposed with various $\begin{pmatrix} C^{(2)} & 0 \\ 0 & C'^{(2)} \end{pmatrix}$ s and block diagonal matrices (with diagonal blocks) for arbitrary $2^n$-by-$2^n$ matrices $A$ and $B$. Therefore, $\begin{pmatrix} A & 0 \\ 0 & B \end{pmatrix}$ also can be decomposed into block diagonal matrices and block circulant matrices. Let us consider $\begin{pmatrix} A' & B' \\ B' & A' \end{pmatrix}$ and $\begin{pmatrix} I & 0 \\ I & I \end{pmatrix}$ where $A' = A + B$ and $B' = A - B$ are another arbitrary $2^n$-by-$2^n$ matrices. By considering the following equalities:

$$\begin{cases} \begin{pmatrix} A' & B' \\ B' & A' \end{pmatrix} = \begin{pmatrix} I & I \\ I & -I \end{pmatrix}\begin{pmatrix} A & 0 \\ 0 & B \end{pmatrix}\begin{pmatrix} I & I \\ I & -I \end{pmatrix} \\ \begin{pmatrix} I & 0 \\ I & I \end{pmatrix} = \begin{pmatrix} \frac{5}{4}I & 0 \\ 0 & \frac{5}{3}I \end{pmatrix}\begin{pmatrix} I & 2I \\ 2I & I \end{pmatrix}\begin{pmatrix} 2I & 0 \\ 0 & I \end{pmatrix}\begin{pmatrix} I & I \\ I & -I \end{pmatrix}\begin{pmatrix} \frac{1}{5}I & 0 \\ 0 & \frac{1}{3}I \end{pmatrix} \end{cases}, \qquad \text{(S7)}$$

we now know that $\begin{pmatrix} A' & B' \\ B' & A' \end{pmatrix}$ and, hence, $\begin{pmatrix} I & 0 \\ I & I \end{pmatrix}$ can be decomposed into block diagonal matrices and block circulant matrices. For arbitrary $2^n$-by-$2^n$ invertible matrix $P$, $\begin{pmatrix} I & 0 \\ P & I \end{pmatrix}$ can be decomposed into block diagonal matrices and block circulant matrices because $\begin{pmatrix} I & 0 \\ P & I \end{pmatrix} = \begin{pmatrix} I & 0 \\ 0 & P \end{pmatrix}\begin{pmatrix} I & 0 \\ I & I \end{pmatrix}\begin{pmatrix} I & 0 \\ 0 & P^{-1} \end{pmatrix}$. Since every matrix is a sum of invertible matrices, $\begin{pmatrix} I & 0 \\ A' & I \end{pmatrix} = \begin{pmatrix} I & 0 \\ P+Q & I \end{pmatrix} = \begin{pmatrix} I & 0 \\ P & I \end{pmatrix}\begin{pmatrix} I & 0 \\ Q & I \end{pmatrix}$ also can be decomposed into block diagonal matrices and block circulant matrices where $P$ and $Q$ are arbitrary $2^n$-by-$2^n$ invertible matrices and $A'$ is an arbitrary $2^n$-by-$2^n$ matrix. Finally, we have $\begin{pmatrix} A & 0 \\ 0 & B \end{pmatrix}$, $\begin{pmatrix} I & 0 \\ A' & I \end{pmatrix}$, $\begin{pmatrix} I & B' \\ 0 & I \end{pmatrix}$ where all three matrices can be decomposed into block diagonal matrices and block circulant matrices for arbitrary $2^n$-by-$2^n$ matrices, $A$, $B$, $A'$, and $B'$. By following the arguments in Ref. 40 in the main text, it can be concluded that an arbitrary $2^{n+1}$-by-$2^{n+1}$ matrix can be decomposed into block diagonal matrices and block circulant matrices of the same size.

It is straightforward that an arbitrary block circulant matrix can be decomposed as $F^{(2)}D^{(2)}F^{(2)}$ where $F^{(2)} = F \otimes [1,\ 0;\ 0,\ 1]$ ($F$ is a $2^n$-by-$2^n$ discrete Fourier transform matrix) and $D^{(2)}$ is a $2^{n+1}$-by-$2^{n+1}$ block diagonal matrix. Therefore, we prove that an arbitrary $2^{n+1}$-by-$2^{n+1}$ matrix can be decomposed into block diagonal matrices and $F^{(2)} = F \otimes [1,\ 0;\ 0,\ 1]$. In the $X \otimes P$ basis, a block diagonal matrix represents a universal metasurface, where each block represents a Jones matrix for a specific spatial point and $F^{(2)}$ represents a conventional lens.

## Supplementary Figures

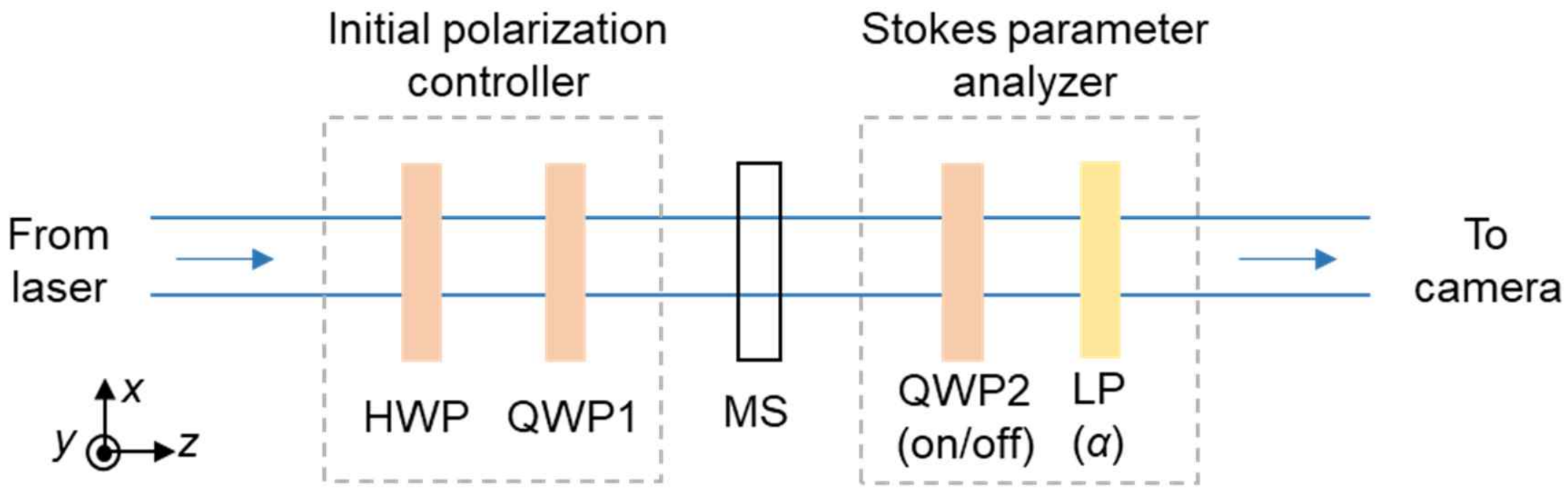

**Fig. S1. Optical setup for full-Stokes polarimetry.** Input polarization can be controlled by the orientation angle of HWP and QWP1 of the initial polarization controller. Stokes parameters can be retrieved from four intensity measurements, for example, $I_1 = I(\text{off}, \alpha = 0°)$, $I_2 = I(\text{off}, \alpha = 90°)$, $I_3 = I(\text{off}, \alpha = 45°)$ and $I_4 = I(\text{on}, \alpha = 45°)$ where on/off represents presence/absence of QWP2 (fast axis is aligned in $y$-axis) and α is orientation angle of LP of Stokes parameter analyzer. Then, $S_0 = I_1+I_2$, $S_1 = I_1+I_2$, $S_2 = 2I_3–I_1–I_2$, and $S_3 = 2I_4–I_1–I_2$. HWP: half-wave plate; QWP: quarter-wave plate; MS: metasurface; LP: linear polarizer.

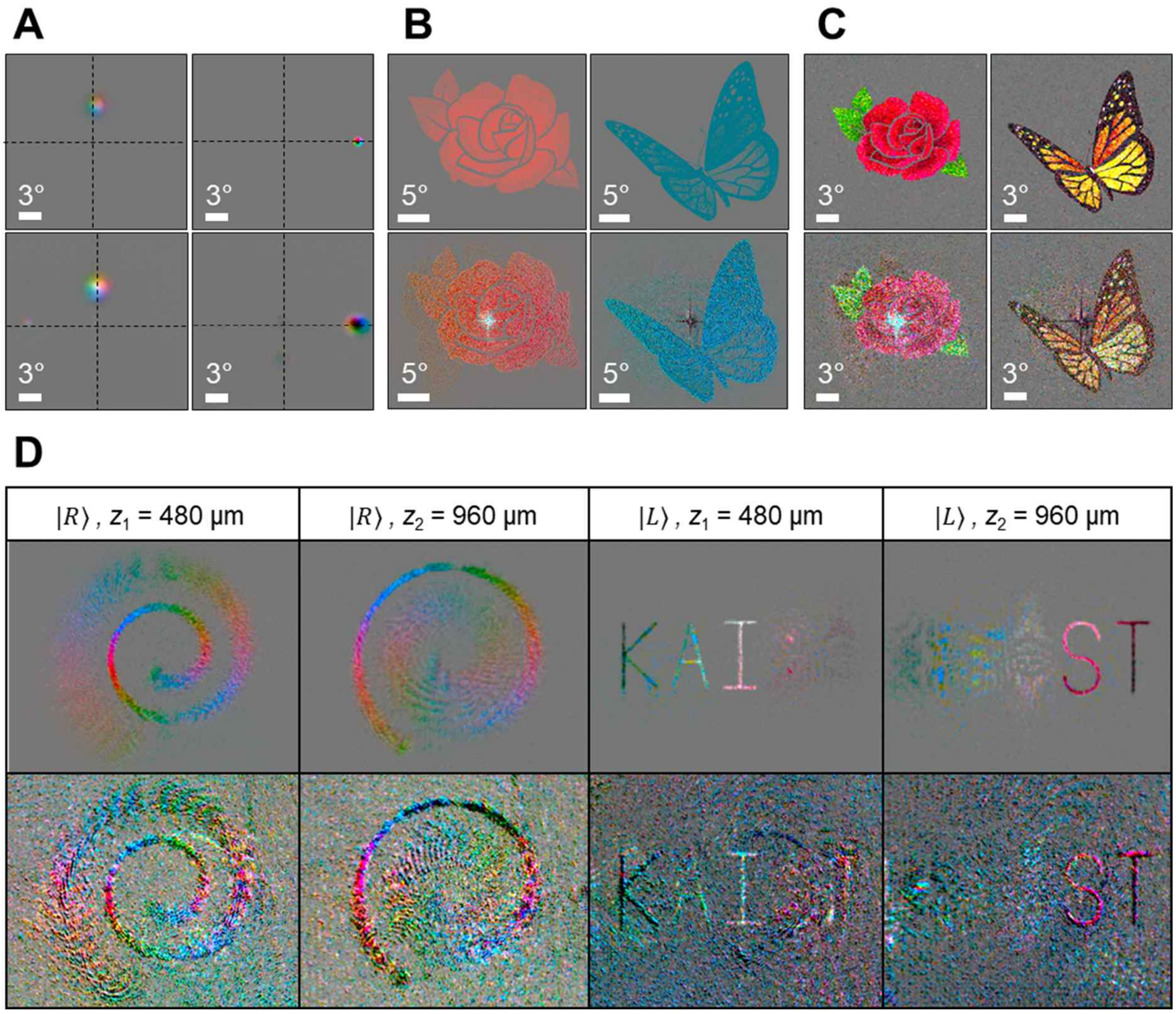

**Fig. S2. Comparison of holographic images from Fig. 3–4 in the main text between numerical calculation and experiment.** (**A**) Two orthogonal Poincaré beams. Throughout the figure, the top (bottom) panels show numerical calculation (measured) results. Left (right) panels show results for RCP (LCP) input polarization state. (**B**) Two independent holographic images where the output polarization state is not the conjugation of the input polarization state. Left (right) panels show results for RCP (LCP) input polarization state. (**C**) Two independent vectorial holographic images. Left (right) panels show results for RCP (LCP) input polarization state. (**D**) Two independent three-dimensional vectorial holographic images. Each column shows results for different input polarization states (RCP or LCP) and different focal plane ($z$ = 480 μm or $z$ = 960 μm).

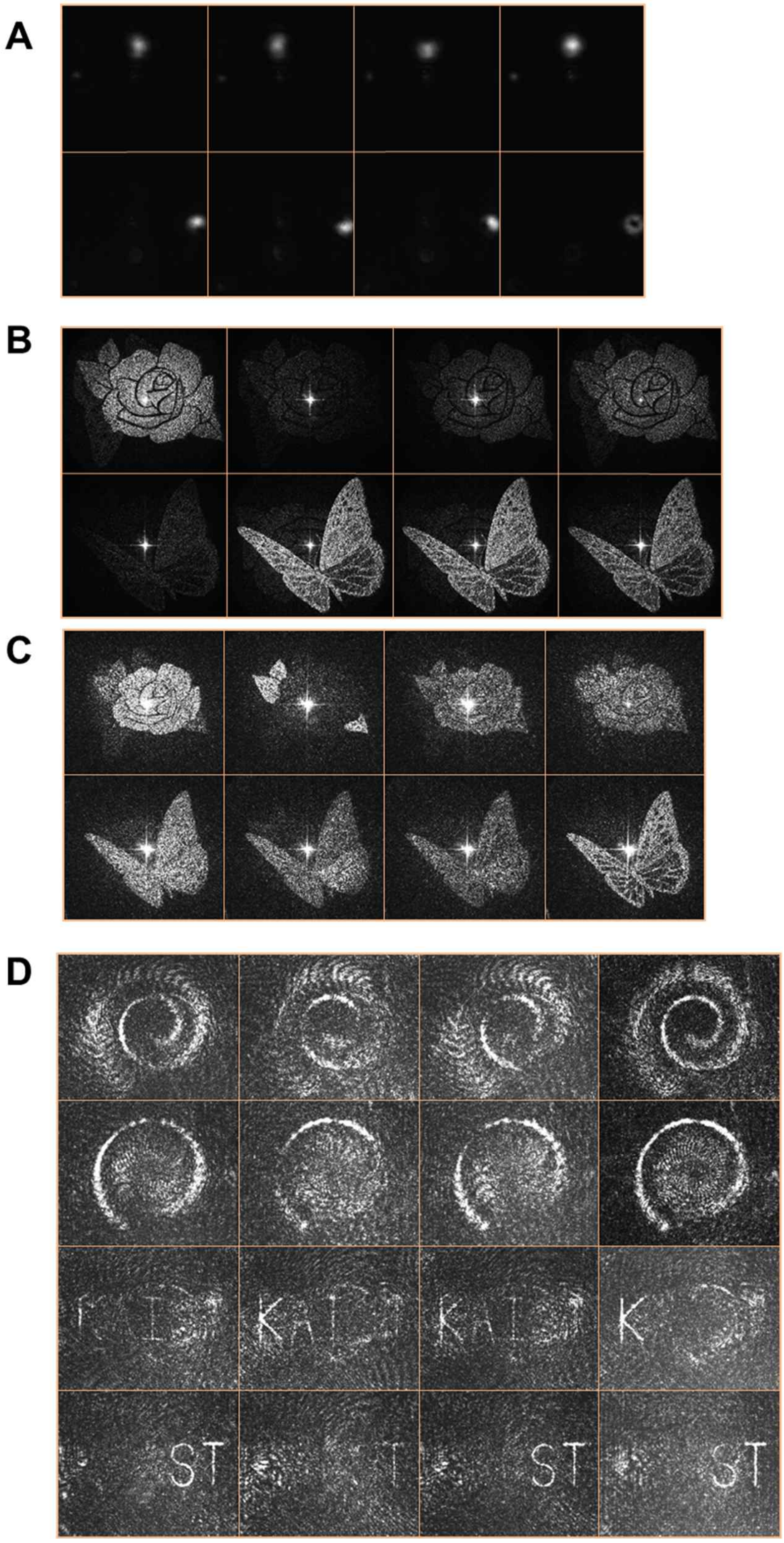
A
B
C
D

**Fig. S3. Raw data from the full-Stokes polarimetry.** Throughout the figure, intensity measurements of $I_1$-$I_4$ are represented in the panels from leftmost to rightmost, respectively. (**A**) Two orthogonal Poincaré beams. Top (bottom) panels show measurement for RCP (LCP) input. (**B**) Two independent holographic images where output polarization is different from the conjugation of input polarization. Top (bottom) panels show measurement for RCP (LCP) input. (**C**) Two independent vectorial holographic images. Top (bottom) panels show measurement for RCP (LCP) input. (**D**) Two three-dimensional vectorial field profiles. From top to bottom, measurement of RCP input at z = 480 μm, RCP input at z = 960 μm, LCP input at z = 480 μm and LCP input at z = 960 μm are represented. The specific configurations of optical setups that vary over the measurements in (A–D), such as the number of mirrors and the orientation of waveplates, are properly considered in order to retrieve quantitative Stokes parameters.

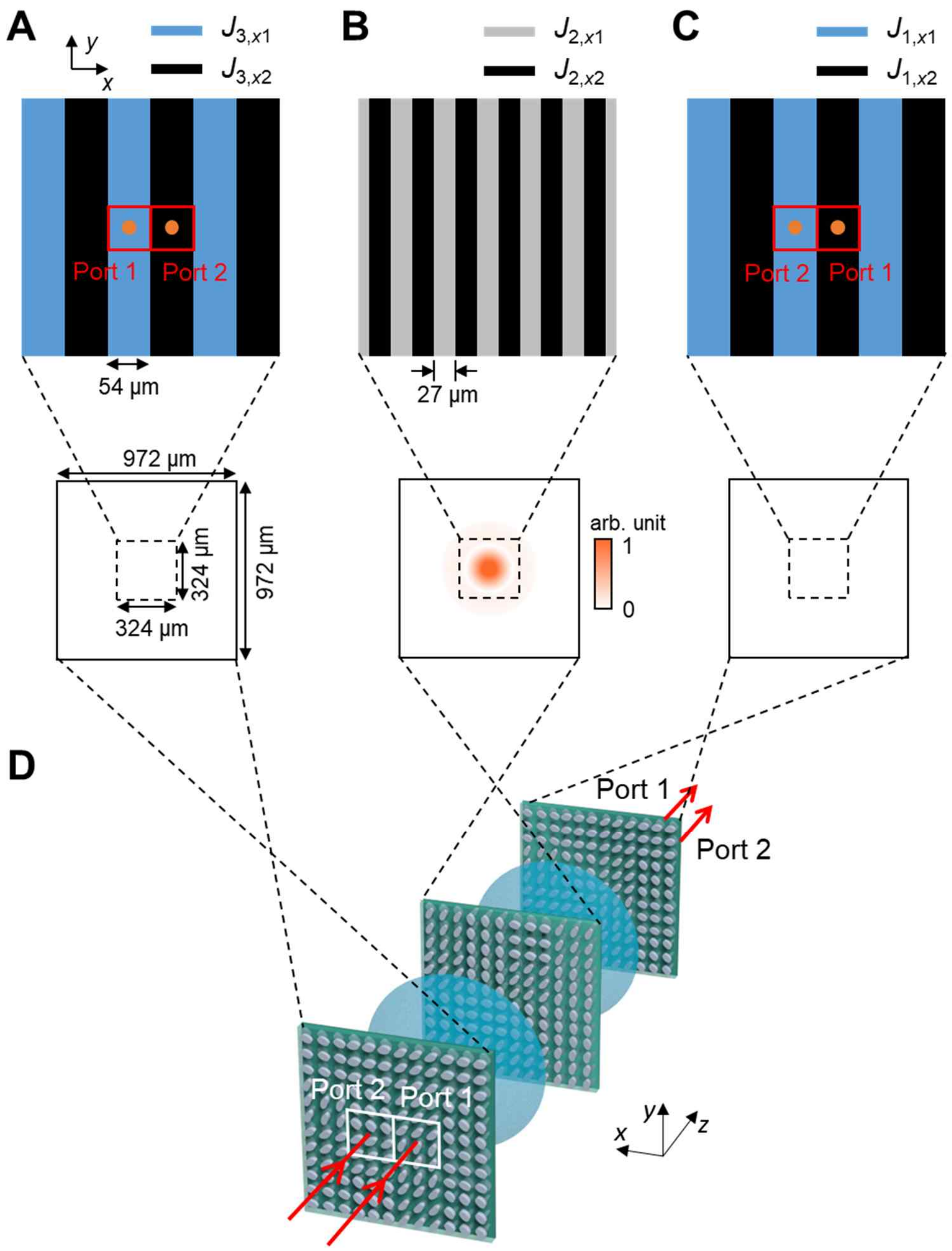

**Fig. S4. Designed Jones matrices of universal metasurfaces for the proposed CNOT gate.** For (**A**–**C**), upper panels show close-up views of universal metasurfaces, and each metasurface (lower panel) is a periodic array of the respective center area. (**A**) Design of the frontend universal metasurface. $D_3^{(2)\prime} = \begin{pmatrix} J_{3,x1} & 0 \\ 0 & J_{3,x2} \end{pmatrix} = \begin{pmatrix} b_1 & 0 & 0 & 0 \\ 0 & b_2 & 0 & 0 \\ 0 & 0 & b_3 & b_4 \\ 0 & 0 & -b_3 & b_4 \end{pmatrix}$ where $b_1$ to $b_4$ are shown in Materials and methods section 9. The ports area is shown with a red rectangular area. Input beam diameter size is shown as an orange dot. Note that $T_q^{-1}$ is omitted in this figure. (**B**) Design of the universal metasurface at the center. $D_2^{(2)\prime} = \begin{pmatrix} J_{2,x1} & 0 \\ 0 & J_{2,x2} \end{pmatrix} = \begin{pmatrix} b_5 & 0 & 0 & 0 \\ 0 & b_6 & 0 & 0 \\ 0 & 0 & b_5 & 0 \\ 0 & 0 & 0 & b_7 \end{pmatrix}$ where $b_5$ to $b_7$ are shown in Materials and methods section 9. The beam amplitude profile under a single port input is shown in the upper panel. It can be confirmed that the transversal device size of around 1 mm$^2$ is large enough to avoid crosstalk in the case of integration. (**C**) Design of the backend universal metasurface. $D_1^{(2)\prime} = \begin{pmatrix} J_{1,x1} & 0 \\ 0 & J_{1,x2} \end{pmatrix} = \begin{pmatrix} b_1 & 0 & 0 & 0 \\ 0 & b_2 & 0 & 0 \\ 0 & 0 & b_3 & -b_3 \\ 0 & 0 & b_4 & b_4 \end{pmatrix}$. The ports area is shown with a red rectangular area. Output beam diameter size is shown as an orange dot. Note that $T_q$ is omitted in this figure. (**D**) A schematic of the proposed CNOT gate (not drawn to scale).